\documentclass[aps,prx,twocolumn,superscriptaddress,english,floatfix, longbibliography]{revtex4-2}
\usepackage{graphicx}
\usepackage{float}
\usepackage{physics}
\usepackage{cancel}
\usepackage{overpic}
\usepackage{enumerate}
\usepackage[colorlinks,citecolor=blue,linkcolor=blue,urlcolor=blue]{hyperref}
\usepackage{textcomp}
\usepackage{amsmath}
\usepackage{amssymb}
\usepackage{pgf}
\usepackage{soul}
\usepackage[normalem]{ulem}
\usepackage{lipsum}
\usepackage{comment}
\usepackage{mathrsfs}
\usepackage{standalone}
\usepackage[english]{babel}
\usepackage{bm}
\usepackage{tikz}
\usepackage{orcidlink}

\renewcommand{\(}{\left(}
\renewcommand{\)}{\right)}
\renewcommand{\[}{\left[}
\renewcommand{\]}{\right]}

\newcommand{\calpha}{\alpha^\ast}
\newcommand{\calphas}{\alpha^{\ast2}}

\let\originalleft\left
\let\originalright\right
\renewcommand{\left}{\mathopen{}\mathclose\bgroup\originalleft}
\renewcommand{\right}{\aftergroup\egroup\originalright}

\newcommand{\vect}[1]{\boldsymbol{#1}}
\renewcommand{\vec}[1]{\vect{#1}}

\makeatletter
\newcommand*\bigcdot{{\color{gray}\mathpalette\bigcdot@{1.}}}
\newcommand*\bigcdot@[2]{\mathbin{\vcenter{\hbox{\scalebox{#2}{$\m@th#1\bullet$}}}}}
\makeatother

\def\rme{{\rm {e}}}
\def\rmi{{\rm {i}}}
\renewcommand{\d}{{\rm {d}}}
\def\dt{{\rm {d}}t}

\def\tr{{\rm{Tr}}}
\newcommand{\abss}[1]{\left|#1\right|^2}
\newcommand{\mean}[1]{\left\langle #1 \right\rangle}

\newcommand{\alphatil}{{\Tilde{\alpha}}}

\newcommand{\rhohat}{\hat{\rho}}

\newcommand{\psitilde}{\widetilde{\ket{\psi}}}

\newcommand{\aaa}{\hat{a}}
\newcommand{\daaa}{\hat{a}^\dagger}
\newcommand{\daaas}{\hat{a}^{\dagger 2}}
\newcommand{\aaas}{\aaa^{2}}

\newcommand{\Hhat}{\hat{H}}

\newcommand{\Lhat}{\hat{L}}

\newcommand{\ohat}{\hat{O}}

\newcommand{\Ohat}{\hat{O}}

\newcommand{\pscr}{{\mathscr{P}}}

\newcommand{\bea}{\begin{equation}\begin{aligned}}
		\newcommand{\eea}{\end{aligned}\end{equation}}
\newcommand{\be}{\begin{equation}}
	\newcommand{\ee}{\end{equation}}

\renewcommand\rme{{{e}}}
\renewcommand\rmi{{{i}}}
\renewcommand{\d}{{{d}}}
\renewcommand\dt{{{d}}t}

\renewcommand{\selectlanguage}[1]{}

\begin{document}

\title{Measurement-induced phase transitions in monitored infinite-range interacting systems}

\author{Anna Delmonte~\orcidlink{0009-0008-9371-6855}}
\thanks{These authors contributed equally to this work}
\affiliation{SISSA, Via Bonomea 265, I-34136 Trieste, Italy}
\author{Zejian Li~\orcidlink{0000-0002-5652-7034}}
\thanks{These authors contributed equally to this work}
\affiliation{The Abdus Salam International Center for Theoretical Physics, Strada Costiera 11, 34151 Trieste, Italy}
\author{Gianluca Passarelli~\orcidlink{0000-0002-3292-0034}}
\affiliation{Dipartimento di Fisica ``E. Pancini", Universit\`a di Napoli ``Federico II'', Monte S. Angelo, I-80126 Napoli, Italy}
\author{Eric Yilun Song}
\affiliation{JILA, NIST, and Department of Physics, University of Colorado, Boulder, CO, USA}

\author{Diego Barberena~\orcidlink{0000-0002-6845-1807}}
\affiliation{T.C.M. Group, Cavendish Laboratory, University of Cambridge, J.J. Thomson Avenue, Cambridge CB3 0HE, UK}

\author{Ana Maria Rey~\orcidlink{0000-0001-7176-9413}}
\affiliation{JILA, NIST, and Department of Physics, University of Colorado, Boulder, CO, USA}
\affiliation{Center for Theory of Quantum Matter, University of Colorado, Boulder, CO, USA}

\author{Rosario Fazio~\orcidlink{0000-0002-7793-179X}}
\affiliation{The Abdus Salam International Center for Theoretical Physics, Strada Costiera 11, 34151 Trieste, Italy}
\affiliation{Dipartimento di Fisica ``E. Pancini", Universit\`a di Napoli ``Federico II'', Monte S. Angelo, I-80126 Napoli, Italy}

\begin{abstract}

A key challenge in observing measurement-induced phase transitions is the mitigation of the post-selection barrier, which causes the reproducibility of specific sequences of measurement readouts—the trajectory—to be exponentially small in system size. Recent studies suggest that some classes of monitored infinite-range systems alleviate this problem by exhibiting a fast saturation of entanglement, resulting in only a polynomial post-selection overhead. This paper answers whether this feature is inherent in infinite-range systems, due to their underlying semiclassical dynamics. We consider three experimentally relevant monitored models: a Tavis-Cummings model, a Superradiance model, and a Bose-Hubbard dimer, each exhibiting non-trivial monitored dynamics. We unveil the occurrence of entanglement phase transitions in these models, showing how the saturation time is strongly affected by bistability regions, which also prevent the mitigation of the post-selection barrier. Finally, we propose experimental realizations of these models, providing a discussion of post selection from an experimental perspective.
\end{abstract}
\maketitle

\section{Introduction}
Synthetic quantum many-body open systems offer a rich playground to explore an amazing variety of out-of-equilibrium phases and phase transitions~\cite{Noh, Hartmann_2016, cirac2012, Marino2022,Fazio2024}.
The last decade has witnessed impressive progress in the available experimental platforms, from arrays of coupled cavities~\cite{Saxena, Underwood,Fitzpatriv} to cold atoms with controlled dissipation~\cite{Lobser,Langen,Smale}, from trapped ions~\cite{Safavi2018,Mei,Kaplan,Cole2021} to  exciton-polariton condensates in cavities~\cite{Castellanos,Pieczarka} and ultracold atoms in optical lattices \cite{Greiner, Gross2017}. This, in turn, stimulated an intense theoretical effort~\cite{Sieberer_2016, Eisert_2015, Camacho_2022, Vasseur_2016, Bernard_2016, Zejian}. Notable examples, in a long list of very interesting developments, of experimental breakthroughs are the realization of the Mott insulating state with photons~\cite{Ma}, the first dissipative transition in a cavity dimer~\cite{Tikan} and in arrays of long-lived Sr dipoles \cite{song2024}. Non-equilibrium is attained in these systems due to the interplay between the unitary dynamics and the damping/decoherence, they are often referred to as driven-dissipative systems. Essentially in all the cases of interest, the dynamics is Markovian and the evolution of the (mixed) state is governed by the Lindblad equation~\cite{breuer2002theory}.

The dynamics of the system is however profoundly different if one is able to continuously monitor its state~\cite{Jacobs, Wiseman}. In this last case, the quantum system is characterized by a wavefunction whose smooth evolution is interrupted at random times by sudden changes that account for the interaction of the system with the environment~\cite{Carmichael,Molmer}. Averaging over all possible quantum trajectories, i.\,e., averaging over all the measurement outcomes, leads to the mixed-state Lindblad evolution. The dynamics of monitored systems has been extensively investigated, for several decades, in few-level systems and under different monitoring~\cite{PhysRevA.44.3340,PhysRevB.60.5737,PhysRevB.63.115403,PhysRevA.65.052304,PhysRevA.35.198, PRXQuantum.5.020304,Albarelli2023,Gal2024,Gleyzes2007}. 

For many years unravelling the Lindblad dynamics of many-body open systems into its quantum trajectories was mainly considered a numerical tool to simulate open systems. Only recently it was found that a monitored many-body quantum system shows a rich phenomenology that is invisible at the level of its average state, the density matrix \cite{Skinner_prx}. Examples of this sort are the measurement-induced phase transitions that have been predicted initially in monitored quantum circuits \cite{Nahum_prxq,Li_Fisher,Li_Fisher2,Romito_Schomerus,Huse,Choi}, separating phases with different entanglement properties. In Ref. \cite{Skinner_prx} the circuit consisted of a sequence of entangling gates intertwined by local projective measurement on a fraction of the qubits. The interplay between the gates, leads to extensive entanglement and the local projections that disentangle the qubits results in a critical phenomenon occurring as a function of the fraction of measured qubits, separating volume from area law. Most importantly, none of the expectation values of the local observables of the systems would be able to detect this transition. Entanglement, like any other non-linear function of the state, has properties that cannot be ascertained from the knowledge of the average density matrix.  

These initial works~\cite{Li_Fisher,Skinner_prx} stimulated an intense theoretical activity scrutinizing many different facets of measurement-induced phases and transitions. In the most common cases, unitary dynamics will generate entanglement and the process of local measurement that on the other side destroys it; hence this phenomenology should be observed in a quite broad range of systems coupled to an environment/measurement apparatus. This interplay so far has been investigated in quantum circuits~\cite{ippoliti2021entanglementphasetransitions,Sierant_2023,sierant2023controlling,Kalsi_Romito_Schomerus_2022,ware2023sharp} and in Hamiltonian systems~\cite{SwingleSYk,BucchholdFreeF,BuchholdDirac,BuchholdFreeB,FujiAshida,Cao,Kells,Turkeshi,lumia2024measurementinduced,paviglianiti2023multipartiteentanglementin,guido2,Tirrito_Santini_Fazio_Collura_2023}. Both projective measurements and/or generalized measurements~\cite{Szyniszewski2019, Gebhart, Doggen, Romito_Schomerus2, JacobsSteck, leung2023theory} performed by an external environment have been considered as well. There is a deep connection between measurement-induced phases and the encoding/decoding properties of a quantum channel~\cite{Gullans}. 

Despite this large theoretical activity, the experimental evidence of MIPTs is much more limited. As of today, only three pioneering experimental works have been able to observe the monitored many-body dynamics. The first quantum simulation was performed with trapped ions~\cite{Monroe_2021}. The scaling close to the critical point has been explored with the IBM quantum processor~\cite{KohIBM,kamakari2024experimental} and with the Google quantum processor for different settings of the quantum circuits~\cite{Google1}. The reason for this difficulty is of fundamental nature and very often is referred to as the {\it post-selection problem} \cite{gullans2020scalableprobesof,Ippoliti}. 

The problem is very easily stated. Monitored dynamics requires evaluating observables (or any non-linear function of the state) for a given sequence of outcomes of the measurements/jumps, a quantum trajectory. This means that in order to perform the quantum average one should be able to reproduce with sufficiently high probability the {\it same} record of outcomes. While this is certainly possible for few-body systems, it becomes exceedingly difficult in a many-body setting as the probability of reproducing the same trajectory becomes exponentially small both with the system size and the time at which the average should be taken. This is the reason why experiments have been limited to moderate system sizes and considerable efforts were required to increase the system dimension.

Even though most probably there is no way to overcome the post-selection barrier under general conditions, some interesting progress has been made to alleviate (or even suppress) it in specific cases. This is for instance the case of quantum-classical approaches that combine measurement outcomes with classical post-processing~\cite{AltmanFisher, Dehghani_2023}. These methods work efficiently when one is able to simulate efficiently the quantum dynamics on a classical computer (i.\,e., in the area-law phase).  A recent experimental demonstration of these approaches has been reported in~\cite{kamakari2024experimental}. Post-selection is also eluded in the semi-classical regime by combining classical simulations with the output from a quantum simulation~\cite{Li_Delmonte_Turkeshi_Fazio_2024}.
For certain classes of circuits, post-selection overhead can be mitigated by resorting to space-time duality~\cite{Ippoliti,Grover,Google1}. 
Furthermore, a series of works ~\cite{PhysRevLett.131.060403,buchhold2022revealing,ravindranath2022entanglementsteeringin,odea2022entanglementandabsorbing,piroli2022trivialityofquantum,sierant2023controlling,ravindranath2023free,allocca2024statistical,lemaire2023separate,Sierant_2023} explored the possibility of using quantum circuits with feedback, where the evolution is conditioned on the measurement registry to render the measurement-induced transition visible in the average dynamics. In this last case, the observation is indirect and relies on the assumption that the transition captured by the density matrix coincides with that in the monitored dynamics. It should be taken into account however that feedback may alter the physical properties of the system and separate transitions may occur in the monitored and average dynamics. Despite these progresses, the post-selection barrier remains a tough challenge, and progress in overcoming it is of fundamental importance to studying monitored dynamics in quantum many-body systems. 
Post-selection-free monitored dynamics was shown to emerge naturally in a fully connected model~\cite{Passarelli} describing a monitored ensemble of atoms driven by a laser field and in the presence of collective decay.
The aim of this work is to further explore regimes where post-selection can be mitigated. We start from Ref.~\cite{Passarelli} where a specific class of infinite-range models was shown to have only a power-law (in system size) overhead associated with post-selection. The question we would like to address is in which extent this post-selection friendly situation is common to infinite-range interacting monitored systems. In this work, we will consider several different infinite-range monitored Hamiltonian systems that are of relevance for experimental implementation.

The paper is organized as follows. In the next Section, Sec.~\ref{secI}, we will introduce the various models we are going to study, briefly recap their properties, define their monitored dynamics, and introduce the quantities (both linear and non-linear in the quantum state) that we are going to use to characterize the transitions. Specifically, we will study i) a driven-dissipative Tavis-Cummings model, also known as collective resonance fluorescence model, presenting a rich phase diagram, ii) a monitored ensemble of atoms driven by a laser field and in the presence of collective decay, extending the results of~\cite{Passarelli} to the case of finite temperature to include both collective absorption and emission, iii) a driven-dissipative Bose-Hubbard dimer undergoing, depending on the regime of the coupling, either a first-order or a second-order phase transition, as a function of the driving. In Section~\ref{SecIII} we will show the results for the different models. We will see that there are generic features that are common to these cases. At the same time, the existence of underlying classical dynamics is not, by itself, sufficient to dictate the properties of monitored systems, especially for what concerns the post-selection barrier. We will see that a crucial point in this respect is the type of transition the system is undergoing as a function of the monitoring strength. All the models discussed here have an immediate experimental realization, in Section ~\ref{SecExp} we propose their implementation addressing also practical aspects of the post-selection problem.  We then move to the conclusions in Section~\ref{SecIV}. Infinite-range models are therefore a useful playground to test the properties of monitored many-body systems without a post-selection barrier.  Some details of our analysis are included in the Appendices.

\section{Infinite-Range Models and their monitored dynamics}
\label{secI}
In this section, we introduce three distinct models that will be then studied in the rest of the paper. Specifically, we will consider: a Tavis-Cummings model describing a collection of spins in a single-mode cavity, see Fig.~\ref{fig_models}(a); a Superradiant model with collective decay and pumping, see Fig.~\ref{fig_models}(b); a Bose-Hubbard dimer consisting of two coupled cavities with Kerr non-linearity, see Fig.~\ref{fig_models}(c),
that is also implementable with multi-level long-lived atoms in an optical cavity. Our goal is to study and understand the dynamics of quantum correlations present in the systems subject to monitored dynamics. 

In the following part of the Section we briefly define the monitored dynamics we are going to consider and, to keep the presentation self-consistent, we recap the main properties of the models.

\subsection{Monitored dynamics}
\label{MD}
To model the monitored dynamics of our system, where we count events associated with the jump operators $\hat L_i$ occurring with rates $\kappa_i$, we can define 
a stochastic time evolution of the state $\ket{\psi(t)}$ as follows~\cite{Molmer, Jacobs, Wiseman, Carmichael}. After an interval $dt$ the state is either evolved with a non-Hermitian Hamiltonian (in units where $\hbar=1$) 
\begin{eqnarray}\label{eqAnna:nojump}
    \ket{\psi(t+dt)}&=&\frac{e^{-idt\hat H_{\mathrm{nh}}}\ket{\psi(t)}}{\norm{e^{-idt\hat H_{\mathrm{nh}}}\ket{\psi(t)}}}\\    
    \hat H_{\mathrm{nh}} &=& \hat{H}-\frac{i}{2}\sum_{i}\kappa_i \hat L_i ^\dag \hat L_i \, , \nonumber
\end{eqnarray}
with probability $1-\mathscr P$, or is discontinuously changed to
\begin{equation}\label{eqAnna:jump}
    \ket{\psi{(t+dt)}} = \frac{\hat L_i \ket{\psi(t)}}{\norm{\hat L_i\ket{\psi(t)}}}\, ,
\end{equation}
with probability $\mathscr P$. The $i^{\mathrm{th}}$ jump operator is selected with probability $\mathscr P_{i}\, / \, \mathscr P$. The latter process, dubbed quantum jump, happens with a probability proportional to the $\kappa_i$ rate and to the density $\expval{\hat L_i^\dag \hat L_i}$:
\begin{equation}
    \mathscr P =dt \, \sum_{i} \, \kappa_i \,\expval{\hat L_i^\dag \hat L_i}{\psi(t)} = \sum_{i} \mathscr P_{i}.
\end{equation}
The final state $\ket{\psi_\gamma(t_f)}$ is uniquely determined by the sequence of the detection/non-detection of events associated with $\hat L_i$ at each time step. This defines the quantum trajectory $\gamma$. 

Averaging the stochastic dynamics defined above over all possible outcomes leads to the Lindblad equation~\cite{Molmer, Jacobs, Wiseman, Carmichael}
\begin{equation}\label{eqAnna:master-generic}
    \dfrac{\d\hat \rho}{\d t} = - i \left[ \hat H, \hat \rho \right] + \sum_i \kappa_i \left( \hat L_i \hat \rho \hat L^\dag_i -\frac{1}{2}\left\{\hat L_i ^\dag\hat L_i,\hat\rho \right\} \right).
\end{equation}
Here, $\rhohat=\overline{\ketbra{\psi_\gamma}{\psi_\gamma}}$ is the average density matrix, where we adopt the short-hand notation
\bea
    \overline{\phantom{|}\bullet\phantom{|}}\equiv \mathbb{E}_\gamma\[\bullet\]
\eea
to denote trajectory-average quantities. Note that in the case of the quantum expectation of an observable (we omit the subscript $\gamma$ when there is no confusion) $\langle\Ohat\rangle\equiv\langle\psi|\Ohat|\psi\rangle=\tr[\Ohat\ketbra{\psi}{\psi}]$, a linear function of the state, the trajectory average gives
\bea
\overline{\langle\Ohat\rangle}=\Tr\[\Ohat\rhohat\]\,,
\eea
which is a quantity that can be deduced from the average state $\rhohat$, and thus from the Lindblad equation~\eqref{eqAnna:master-generic}. On the other hand, the trajectory average of nonlinear quantities, such as entanglement (as we will be investigating in this work), does not follow from the Lindblad dynamics, as they require the additional information present in the ensemble of trajectories as a direct consequence of monitoring, that is washed out in the average density matrix $\rhohat$.

In Appendix~\ref{app:binned_ev} we propose an alternative conditional master equation that takes into account finite-time-resolution effects that emerge in experimental realizations of the dynamics.

\subsection{Driven-dissipative Tavis-Cummings model}\label{sec:ddtc}

The driven-dissipative Tavis-Cummings model, also known as collective resonance fluoresce model~\cite{Dicke,Baumann_2010,Brennecke_2013,PhysRevLett.121.040503,TavisCummings,TavisCummings1,TavisCummings2, song2024} serves as a natural playground to study the infinite-range interaction between a single-mode cavity and resonant spins. In particular, we are interested in analyzing the steady state of this model, whose richness stems from the dynamic interplay between an external drive, light and matter interaction, and dissipation processes.

\begin{figure*}
    \centering

    \includegraphics[width=\linewidth]{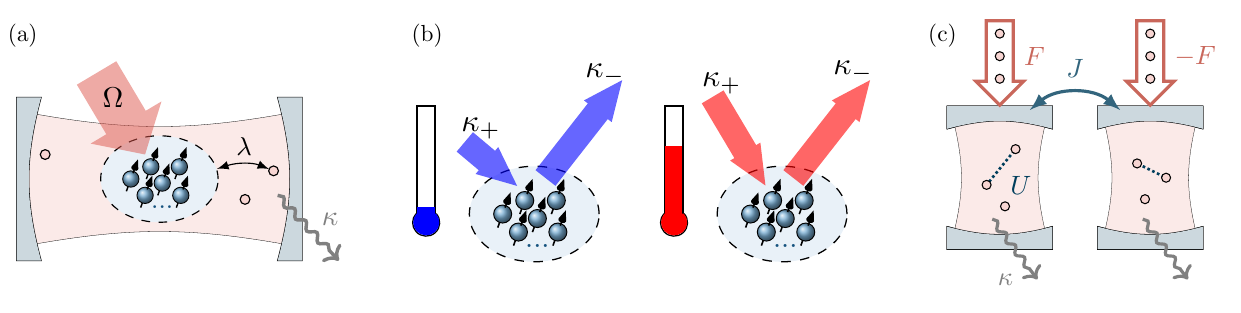}

    \caption{Sketch of the infinite-range models considered in this work. (a) Driven-dissipative Tavis-Cummings Model. Spins are depicted as blue dots with arrows inside a cavity. They interact with photons (red dots) with strength $\lambda$ and are driven by an external field with strength $\Omega$. The photons can leak from the cavity with fixed rate $\kappa$. (b) Superradiant model at finite temperature. Lower temperatures favor the dissipation process with rate $\kappa_-$. Higher temperatures make the opposite process with rate $\kappa_+$ more relevant. (c) Bose-Hubbard dimer and configuration considered in our study. Two photonic cavities are coupled linearly via photon hopping (with rate $J$) and each cavity is nonlinear due to the on-site photon-photon interaction energy $U$. The drive with amplitude $F$ and opposite phases addresses the antibonding mode of the dimer, and each cavity has a dissipation rate of $\kappa$.}
    \label{fig_models}
\end{figure*}

The model, sketched in Fig.~\ref{fig_models}(a), is described  by $N$ spins in a single-mode cavity. The spins and the cavity mode are resonant with characteristic frequency $\omega_c$. The spins collectively interact with the single bosonic mode with a strength $\lambda$  proportional to the single photon Rabi frequency and experience a coherent driving field of strength $\Omega$ and frequency $\omega_d$. The Hamiltonian can be written as $\hat H = \frac{\omega_c}{2} \hat S^z + \omega_c \hat a^\dag \hat a + \frac{\lambda}{\sqrt{N}}(\,\hat a^\dag \hat S^- + \hat a \hat S^+\,) + \Omega (\, \hat S^+ e^{-i\omega_d t} +\hat S^- e^{i\omega_d t} \, )$, where $\hat a^\dag$, $\hat a$ are photons creation and annihilation operators, and $\hat S^{\mu}=1/2\, \sum_{i=1}^N \hat \sigma_i^{\mu}$ with $\mu = x,y,z,\pm$, are collective spin operators.

In the frame rotating at the driving frequency, and assuming the drive to be resonant with the cavity, $\omega_c=\omega_d$,  the Hamiltonian reduces to the simpler form
\begin{equation} \label{eqAnna:Ham_Dicke}
    \hat H = \Omega (\hat S^+ + \hat S^- ) +\frac{\lambda}{\sqrt{N}} \left( \hat a^\dag \hat S^- + \hat a \hat S^+\right)\,.
\end{equation}

Infinite range interactions allow us to express the Hamiltonian through $\hat S^\mu$ spin operators, highlighting the collective nature of the model.
Due to $S^2$ conservation, starting from initial states with maximal $\hat S^2$ eigenvalues, we can effectively describe this model as a single $N/2$ spin interacting with photons in the cavity.

We monitor the photons leaking out of the cavity with a rate $\kappa$. The jump operator in this case is only one:
    $\hat L = \hat a$.
Using this jump operator, one can derive the stochastic evolution of the state given by equations~\eqref{eqAnna:nojump} and~\eqref{eqAnna:jump}, as well as the Lindblad master equation~\eqref{eqAnna:master-generic} for the average state $\rhohat$:
\begin{equation}\label{eqAnna:MasterEq_Dicke}
    \dfrac{d\hat \rho}{dt} = - i \left[ \hat H, \hat \rho \right] + \kappa \left( \hat a \hat \rho \hat a^\dag -\frac{1}{2}\left\{\hat a ^\dag\hat a,\hat\rho \right\} \right)\,.
\end{equation}

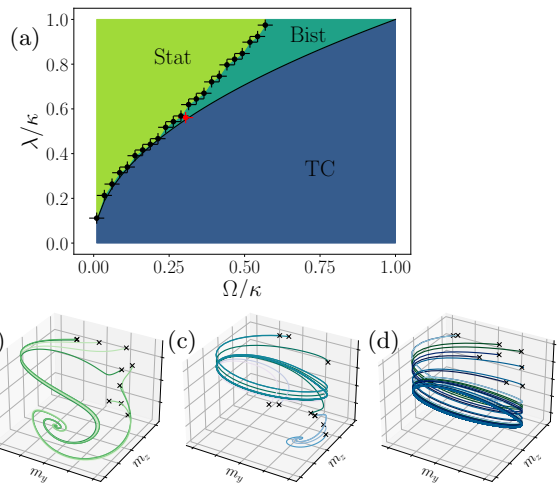
\begin{figure}
    \centering
    \begin{overpic}
        [width=.45\textwidth]{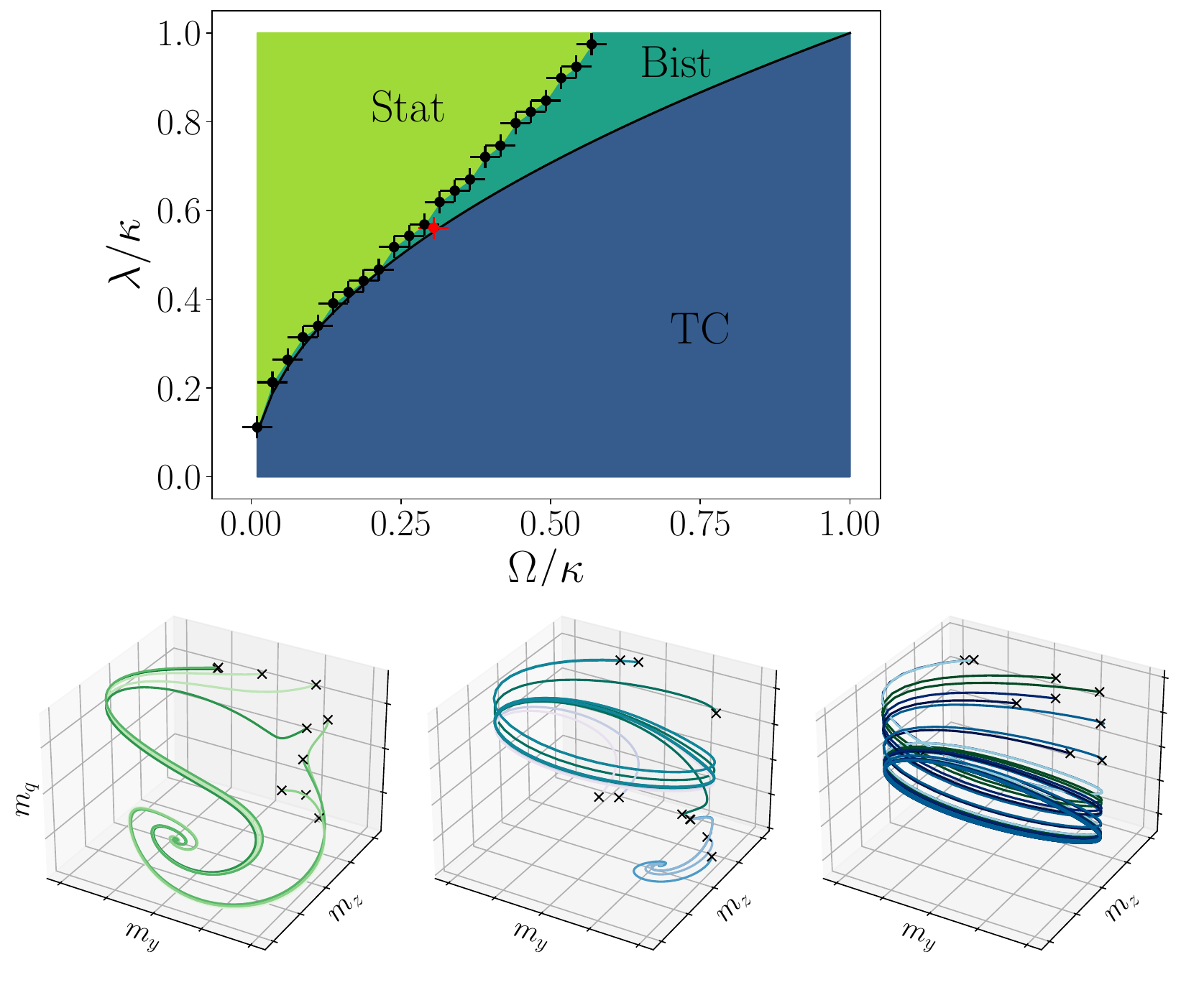} \put (1,25) {(b)} \put (33,25) {(c)} \put (66,25) {(d)} \put (7,75) {(a)}
    \end{overpic}
    \caption{(a) Mean-field phase diagram of the driven-dissipative Tavis Cummings Model, also known as the collective resonance fluorescence model. The border of the time-crystal phase (TC) is known analytically, and the border between the stationary (Stat) and bistability phase (Bist) is determined numerically. The red dot signals the tricritical point. (b), (c), and (d) show semiclassical trajectories in the $(m_y,m_z,m_q)$ space from several initial conditions (signaled by $\times$) respectively in the stationary (b), bistable (c), and time crystal(d) phases.}
    \label{figAnna_2}
\end{figure}

The Heisenberg equations for the magnetizations $m_{\mu}=2\expval{\hat S^{\mu}}/N$ and the quadrature operators $m_{q}=\expval{\hat a^\dag + \hat a}/\sqrt{2N}$ and $m_{p}=i \expval{\hat a^\dag -\hat a}/\sqrt{2N}$
have already been studied in the thermodynamic limit $N\to\infty$~\cite{Mattes}. The steady-state diagram has three phases as in Fig.~\ref{figAnna_2}:  a stationary phase corresponding to a fixed point in the evolution [panel (b)]; a time-crystal phase~\cite{Iemini, Piccitto, Prazeres, Tucker2018} corresponding to limit cycles in the magnetization [panel (d)]; a bistable phase characterized by both aforementioned behaviors (depending on the choice of the initial state) [panel (b)]. The intermediate, bistable phase, vanishes at the tricritical point $(\lambda^T/\kappa,\Omega^T/\kappa)$, corresponding to the red dot in Fig. ~\ref{figAnna_2}. For $\lambda/\kappa<\lambda^T/\kappa$, the system undergoes a continuous phase transition from the stationary phase to the time-crystal phase at $\Omega_C/\kappa=\lambda^2/\kappa^2$. With increasing $\Omega$, at fixed $\lambda/\kappa\ge\lambda^T/\kappa$, the system undergoes a discontinuous phase transition from stationary to time crystal behavior at a critical point (in the bistable phase) dependent on the initial state. For more details on the mean-field limit of the model, see the Appendix~\ref{app:TavisCummings}.

\subsection{Driven spins with collective absorption and emission}\label{sec:dscae}

We will also study the correlation dynamics of a system of $N$ noninteracting two-level atoms, represented in Fig.~\ref{fig_models}(b). The Hamiltonian is given by
\begin{equation}
    \hat H = \frac{1}{2}\Omega \sum_i \hat \sigma^x_i = \Omega \hat S_x,
\end{equation}
with $\Omega$ being the Rabi frequency.

We monitor the system through the two jump operators $\hat L_1 = \hat S_-$ and $\hat L_2 = \hat S_+$.
The dynamics of the average state $\hat \rho$ is then described by the Lindblad equation~\eqref{eqAnna:master-generic},
\begin{equation}\label{eq:btc-finite-t-lindblad}
    \frac{d\hat\rho}{dt} = -i [\hat H, \hat \rho] + \kappa_- \mathcal{D}_-[\hat \rho] + \kappa_+ \mathcal{D}_+[\hat \rho],
\end{equation}
 where the two dissipators $\mathcal{D}_\alpha$ (with $\alpha = \pm$) read
\begin{equation}\label{eq:btc-finite-t-dissipators}
    \mathcal{D}_\alpha[\hat \rho] = \hat S_\alpha \hat \rho \hat S_\alpha^\dagger - \frac{1}{2} \lbrace \hat S_\alpha^\dagger \hat S_\alpha, \hat \rho \rbrace,
\end{equation}
and $\kappa_\alpha$ is their corresponding rate. This spin model can be effectively obtained through a coupling with the infinite-range mode of a finite-temperature electromagnetic cavity. In this framework, the cavity is responsible for two competing effects on the dynamics of the atomic ensemble: the jump operator $\hat S_-$ pushes the total magnetization $\vec{m} = \vec{S} / S$ (with $S = N /2$) towards the south pole of the generalized Bloch sphere, whereas $\hat S_+$ does the opposite and pushes the system towards the north pole. We can thus regard the south pole as an effective zero-temperature state~\footnote{Nothing changes if we swap the roles of the north and south poles.} and define the effective bath temperature $T = 1/\beta$ through the detailed balance condition 
\begin{equation}\label{eq:btc-finite-t-detailed-balance}
    \kappa_+ / \kappa_- \equiv e^{-\beta \Omega_B},
\end{equation}
where $\Omega_B$ is the typical energy of the bath (we will consider $\Omega_B = \kappa$ for simplicity). The detailed balance condition is satisfied by choosing 
\begin{equation}\label{eq:btc-finite-t-rates}
    \kappa_- = \frac{\kappa (1 + n_\beta)}{S}, \qquad \kappa_+ = \frac{\kappa n_\beta}{S},
\end{equation}
where $n_\beta = 1/[\exp(\beta \Omega_B) - 1]$ is the Bose-Einstein distribution of the photons in the cavity and $\kappa$ is the energy scale of the dissipative part of the dynamics. The picture is clear: if the bosonic mode is empty ($n_\beta = 0$, at zero temperature), the bath cannot provide energy to the ensemble, which can only relax towards the south pole. Instead, if $n_\beta > 0$, the bath can give energy to the ensemble and push it up the Dicke ladder~\cite{breuer2002theory}.

The superradiant model proposed in Ref.~\cite{Iemini}, whose entanglement properties were later analyzed~\cite{Passarelli}, corresponds to the zero-temperature limit of this model, which also features a time-crystal (TC) phase in the weakly dissipative regime. This finite-temperature extension was partly analyzed in Ref.~\cite{carollo2023}, where the authors studied the quantum thermodynamics of a dissipative time crystal. 
The mean-field (MF) equations for the evolution of the magnetizations in this model do not depend on temperature: the dissipative phase transition between the TC phase and the symmetry-unbroken (finite-$m_z$) phase exists for all $T$ at $\Omega = \kappa$ and the phase diagram coincides with the zero-temperature one. More details about the mean-field solution are in Appendix~\ref{app:FiniteTSpin}. Despite the fact that the mean-field equations (and, in general, all observables in the thermodynamic limit) do not depend on $T$, we will show in Sec.~\ref{res_finiteTSpin} that the bipartite half-system entanglement entropy $S_{N/2}$ of an ensemble of quantum trajectories is instead dependent on $T$, in the magnetized phase and at the critical point, though its scaling with the system size generally remains the same except at the critical point.

\subsection{Driven-dissipative Bose-Hubbard Dimer}\label{sec:bhd}
Another infinite-range model that we will investigate in this study is the driven-dissipative Bose-Hubbard dimer, which is a semiclassical model yet with rich entanglement dynamics. As sketched in Fig.~\ref{fig_models}(c), the model describes two nonlinear photonic cavities coupled via photon hopping and subjected to one-photon drive and dissipation. The Bose-Hubbard dimer has been extensively studied and is known to present a complex phase diagram~\cite{casteelsOpticallyBistableDrivendissipative2017,lledoDrivenBoseHubbardDimer2019,pudlikDynamicsEntanglementDissipative2013,casteelsQuantumEntanglementSpatialsymmetrybreaking2017,giraldoSemiclassicalBifurcationsQuantum2022}. In this work, we study the entanglement dynamics of the monitored Bose-Hubbard dimer and investigate the entanglement phase transitions in different regimes.

The system is described by the following Hamiltonian (in the frame rotating at the drive frequency),
\bea\label{eq:bh-ham}
    \Hhat &= \sum_{i=1}^{2}\left[ -\Delta\daaa_i\aaa_i + \dfrac{U}{2}\daaas_i\aaas_i+F_i(\daaa_i+\aaa_i) \right] \\&- J(\daaa_1\aaa_2+\daaa_2\aaa_1)\,,
\eea
where $\Delta\equiv \omega_d-\omega_c$ is the detuning between the drive frequency ($\omega_d$) and the bare cavity frequency ($\omega_c$), $U$ is the on-site photon interaction strength, $F_i$ is the driving amplitude on each site and $J$ is the photon-hopping coupling strength between the two sites.
We subject the dimer to local dissipation channels with Lindblad jump operators $\Lhat_{1,2}=\aaa_{1,2}$ with rate $\kappa$,  corresponding to the monitoring of photons being emitted from the two cavities. The dynamics for the average state $\rhohat$ is given by the master equation~\eqref{eqAnna:master-generic},
\bea\label{eq:bh-lindblad}
\dfrac{\d\rhohat}{\d t} = -\rmi[\Hhat,\rhohat] + \kappa\sum_i\(\aaa_i\rhohat\daaa_i-\dfrac{1}{2}\{\daaa_i\aaa_i,\rhohat\}\)\,.
\eea
This system admits a well-defined thermodynamic limit under the scaling
\bea\label{eq:bh-scaling}
    U\longrightarrow \tilde{U}/N,\quad F_i\longrightarrow \sqrt{N}\tilde{F}_i\,,
\eea    
which can be seen as the limit of having infinitely many photons as $N\to\infty$ (cf. Appendix~\ref{app:BHDimer}). This scaling also allows us to identify each cavity mode as the $k=0$ momentum mode of a Bose-Hubbard lattice with $N$ sites~\cite{casteelsCriticalDynamicalProperties2017}. Therefore, the Bose-Hubbard dimer can be identified with a photonic lattice with highly non-local coupling and dissipation and is of infinite-range nature by construction. A parallel visualization can be done using multilevel atoms in cavities. A discussion of this case is provided in the experimental implementation section below \ref{DBH}.
The linear part of the Hamiltonian~\eqref{eq:bh-ham} admits two normal modes, i.\,e., the bonding mode $\aaa_+\equiv (\aaa_1+\aaa_2)/\sqrt{2}$ with eigen frequency $\omega_+=-\Delta-J$, and the antibonding mode $\aaa_-\equiv (\aaa_1-\aaa_2)/\sqrt{2}$, with eigenfrequency $\omega_-=-\Delta+J$. In the rest of this discussion, we will consider a driving profile of $F_1=-F_2\equiv F$, which is driving the antibonding mode $\hat{a}_{-}$ of the dimer. We also consider a detuning $\Delta$ that is sufficiently redshifted from the higher-energy mode but sufficiently above the lower-energy mode. The order in energy of the two modes will depend on the sign of the coupling constant $J$:
\begin{itemize}
    \item $J>0$: The driven mode $\hat{a}_{-}$ is higher in energy. This regime admits a second-order phase transition when the drive strength $F$ is tuned, corresponding to the spontaneous breaking of the spatial symmetry of the dimer~\cite{casteelsQuantumEntanglementSpatialsymmetrybreaking2017}.
    \item $J<0$: The driven mode $\hat{a}_{-}$ is lower in energy. This regime is dominated by the first-order phase transition corresponding to the usual optical bistability of the lower mode~\cite{garbinSpontaneousSymmetryBreaking2022}. Note that the negative hopping has been experimentally realized in photonic crystals~\cite{haddadiPhotonicMoleculesTailoring2014,hamelSpontaneousMirrorsymmetryBreaking2015} and can also be achieved in circuit QED via reservoir engineering~\cite{liDissipationinducedAntiferromagneticlikeFrustration2021}; otherwise, a completely equivalent setup can be obtained by setting $J>0$ and $F_1=F_2$, which is still driving the lower-energy mode.
\end{itemize}

We will study the Bose-Hubbard dimer in these two different regimes using the Gaussian trajectory approximation (cf. Appendix~\ref{sec:gaussian-bhd}) and discuss the different behavior exhibited by the entanglement dynamics in Sec.~\ref{sec:result-bose-hubbard}.

\section{Results}
\label{SecIII}

The models listed above were extensively studied in the past 
and the phase diagram associated with the average state (steady-state of the Lindblad equation) is known. Here we will present our results for the steady state of the entanglement entropy of the monitored system. The main questions we will address are related to i) the (possible) existence of measurement-induced phase transition, ii) their location (their occurrence may/may not coincident with the phase diagram dictated by the Lindblad equation), and iii) the possibility to observe them (whose difficulty is ultimately related to the post-selection barrier). More specifically on this last point we would like to understand whether the semi-classical nature of our models mitigates the post-selection barrier. Notice however that all the properties we study are related to the entanglement entropy and may, in general, vary depending on the non-linear quantity chosen to study the monitored dynamics. 

To tackle the analysis of quantum correlations, we thoroughly study the dynamics and extract the time needed to reach the steady state, which we refer to as the saturation time.
We will see that all the above-listed models have a transition in the entanglement when monitored. The identification and analysis of the properties of these steady-state phase transitions allow for a direct comparison with the transition detected by the average state (visible in observables like magnetization, density,...).

\subsection{Driven-dissipative Tavis-Cummings model}
\label{res_TC}

In this section, we characterize the quantum correlations in the monitored driven-dissipative Tavis-Cummmings Model, focusing on the correlations between spins and intra-cavity photons. The trajectory-averaged  entanglement entropy is defined as
\begin{equation}
    \overline S_E(t) = -\frac{1}{N_\mathrm{traj}}\sum_{\gamma=1}^{N_\mathrm{traj}} \tr\left[\hat\rho_\mathrm{spins}^{(\gamma)}\log\hat\rho_\mathrm{spins}^{(\gamma)}\right]\, ,
\end{equation}
with $\hat\rho_\mathrm{spins}^{(\gamma)}=\Tr_\mathrm{cav}\left[\ket{\psi_\gamma}\bra{\psi_\gamma}\right]$ the reduced density matrix of the spins. The average of this quantity over the time window $\kappa t\in [0,+\infty)$, labeled as $\tilde S_E$, will also be useful in our work since it can capture the steady state properties of quantum correlations.

To have a full understanding of the entanglement entropy behavior across the parameters space $(\Omega/\kappa,\lambda/\kappa)$, we will need to refer to some properties of the magnetization, specifically we will study in some cases the trajectory-averaged magnetization $\overline m_z$, and its time-average $\tilde m_z$ computed over the time window $\kappa t\in [0,+\infty)$.

Before analyzing the result, some details on the simulations are needed. The unraveling of the Lindblad equation has been carried out using the QuTiP package~\cite{Qutip,Qutip2}. The quantum jump unraveling has already been applied to several spin semiclassical systems with $\hat S^2$ conservation~\cite{Passarelli, Cabot}, whose Hilbert space $\mathcal H$ has dimension scaling linearly with $N$, the number of spins. A difference in this cavity model stands out: the bosonic Hilbert space does not have a finite size and needs to be truncated to allow for numerical simulations. To make sure that the truncation does not affect the dynamics of the system, one can check the probability of occupation of the $M^\text{th}$ Fock state: the truncation to the $M_{\mathrm{max}}$-th Fock state can be chosen if $P_M(t)=\expval{\hat\Pi_{M_{\mathrm{max}}}}{\psi(t)}\le \epsilon$ with $\hat\Pi_{M_{\mathrm{max}}}=\ket{M_{\mathrm{max}}}\bra{M_{\mathrm{max}}}$, $M_{\mathrm{max}}\in\mathbb N$. Using this method with $\epsilon=10^{-5}$, one finds that the size of the full Hilbert space is
\begin{equation}
    \mathrm{dim}\left\{\mathcal H\right\} = (M_{\mathrm{max}}+1)(N+1) =\chi N(N+1)\, ,
\end{equation}
with $\chi$ depending on $\lambda/\kappa$ and $\Omega/\kappa$. In general $\chi \lesssim 1$ below the tricritical point, while $\chi\!>\! 1$ above the tricritical point.\\

In Sec.~\ref{lambda02} we focus on the regime 
$\lambda/\kappa=0.2$; in Sec.~\ref{lambda08} we focus on $\lambda/\kappa=0.8$.

\subsubsection{Quantum correlations at $\lambda/\kappa=0.2$}\label{lambda02}


We start the study of the entanglement entropy by reminding some features of the magnetization transition that will be useful as a reference for the following discussion.
The mean-field phase diagram of the magnetization indicates that the region $\lambda/\kappa=0.2$ hosts a continuous phase transition from stationary phase to time-crystal phase at $\Omega_C/\kappa=0.04$. Moreover, the steady state is unique and this region shows no dependence on the choice of initial state.

We carry out the simulations of the entanglement setting as initial state $\ket{\psi_0}=\ket{\psi_\mathrm{cav}}\otimes\ket{\psi_{\mathrm{spin}}}=\ket{0}\otimes\ket{\uparrow,\uparrow,...,\uparrow}$ (any other state belonging to the same symmetry sector would lead to the same conclusions).

\begin{figure}
    \centering    {\begin{overpic}
    [width=0.48\linewidth]{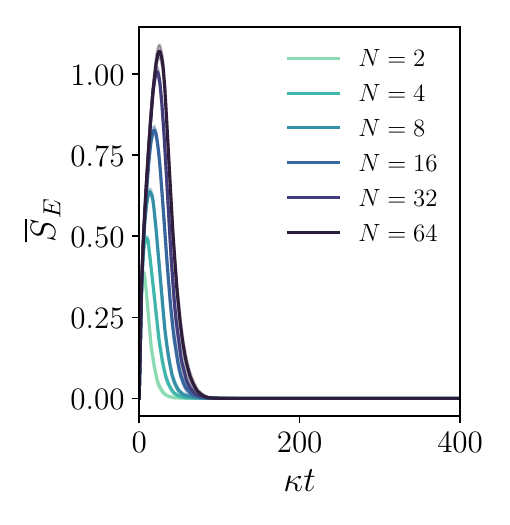}
    \put (0, 85) {(a)}
    \end{overpic}}
    {\begin{overpic}
    [width=0.48\linewidth]{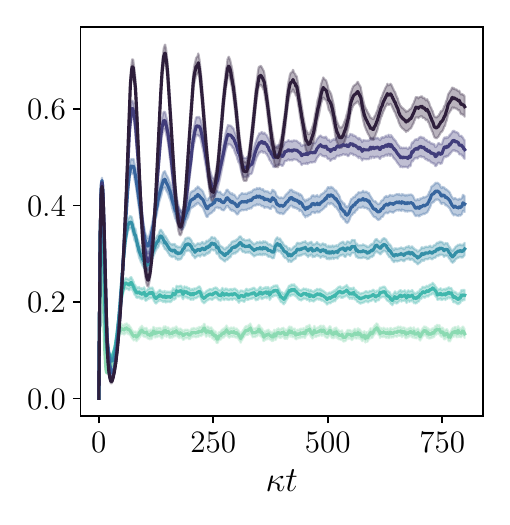}
    \put (-5, 85) {(b)}
    \end{overpic}}
    \caption{Dynamics of the trajectory-averaged entanglement entropy in the (a) stationary phase $\Omega/\kappa=0.01$ and (b) in the time-crystal phase $\Omega/\kappa=0.06$.  All the pictures are obtained for $\lambda/\kappa=0.2$ and from $N_{\mathrm{traj}}=5000$ trajectories. The legend is shared.} 
    \label{figAnna_4}
\end{figure}

\begin{figure*}
    \centering
    {\begin{overpic}
    [width=0.47\linewidth]{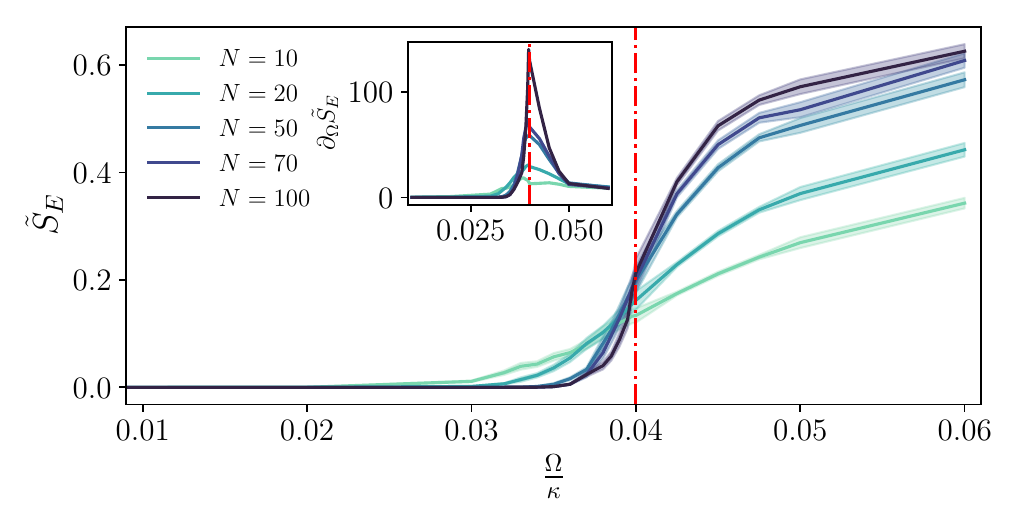} \put (0,44) {(a)}   
    \end{overpic}}
    {\begin{overpic}
    [width=0.25\linewidth]{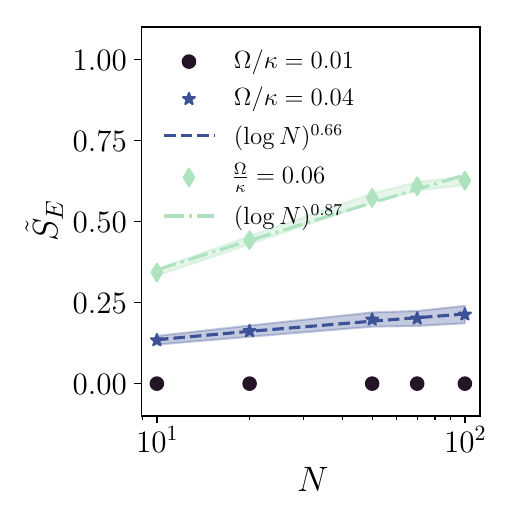} \put (0, 85) {(b)}
    \end{overpic}}
    {\begin{overpic}
    [width=0.25\linewidth]{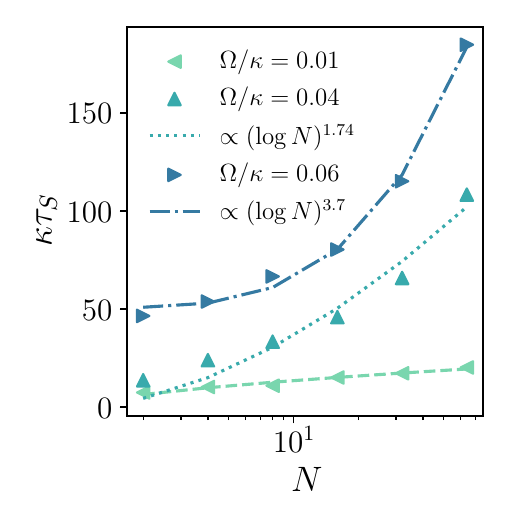} \put (0, 85) {(c)}
    \end{overpic}}
    \caption{(a) $\tilde S_E$ for different sizes and its derivative in the inset. The vertical dash-dot line signals the maximum of the derivative and corresponds to the critical point for the mean-field magnetization. (b) Scaling of $\tilde S_E$ with the size of the system. (c) Saturation time $\tau_S$ with respect to $N$ for different values of $\Omega/\kappa$}.  
    \label{figAnna_5}
\end{figure*}

Fig.~\ref{figAnna_4} shows the time-evolution of entanglement entropy between photons and spins averaged over trajectories, $\overline{S_E}$.  From these plots, we notice that the qualitative features of the temporal behavior of the magnetization is present also in the entanglement dynamics. The first plot corresponds to $\Omega/\kappa=0.01$ which would yield a stationary behavior for the magnetization. Similarly, the entanglement entropy has a stationary behavior. The same holds for the second plot ($\Omega/\kappa=0.06$): for this choice of $\Omega/\kappa$ the magnetization has oscillations and so does the entanglement entropy.

The time-average of $\tilde S_E$ carries important information about the steady state properties of the entanglement entropy. In Fig.~\ref{figAnna_5}(a) this quantity exhibits a phase transition between a non-entangled region in which
$\tilde S_E\propto N^0$ and a highly-entangled one with $\tilde S_E\propto \log^p N$ 
 with $p$ dependent on $\Omega$, $p<1$. The red dot-dashed line at $\Omega_C/\kappa=0.04$ signals the maximum of the of $\partial_\Omega \tilde S_E$
as shown in the inset. We identify this value of $\Omega/\kappa$ as critical point. The entanglement phase transition occurs along with the magnetization one, signaling the entangled nature of the time-crystal phase in the Tavis-Cummings model \cite{Mattes}. The scaling with the size of the entanglement entropy in the two phases is shown in [panel (b)].   

This result integrates what is already found in~\cite{Passarelli}. By adiabatic elimination of the cavity, our Lindblad dynamics can be recast in the master equation of Ref.~\cite{Passarelli} in which a spin-spin entanglement transition has been found in correspondence with the transition between a time crystal behavior and a stationary behavior.

The saturation time obtained from the dynamics (more details in Appendix~\ref{app:sat}), grows as powers of $\log(N)$ yielding a fast-saturating process, see Fig.~\ref{figAnna_5}(c). This makes calculating entanglement entropy with trajectories a partially post-selection-free problem as the probability of measuring a specific trajectory is suppressed only as a polynomial in the system size $N$.

In Appendix~\ref{app:binned_ev} we discuss variations in the scaling that may emerge from experiments due to the finite-time resolutions of the monitoring devices.

\subsubsection{Quantum correlations at $\lambda/\kappa=0.8$}\label{lambda08}
\begin{figure}
    \centering
    {\begin{overpic}
    [width=1\linewidth]{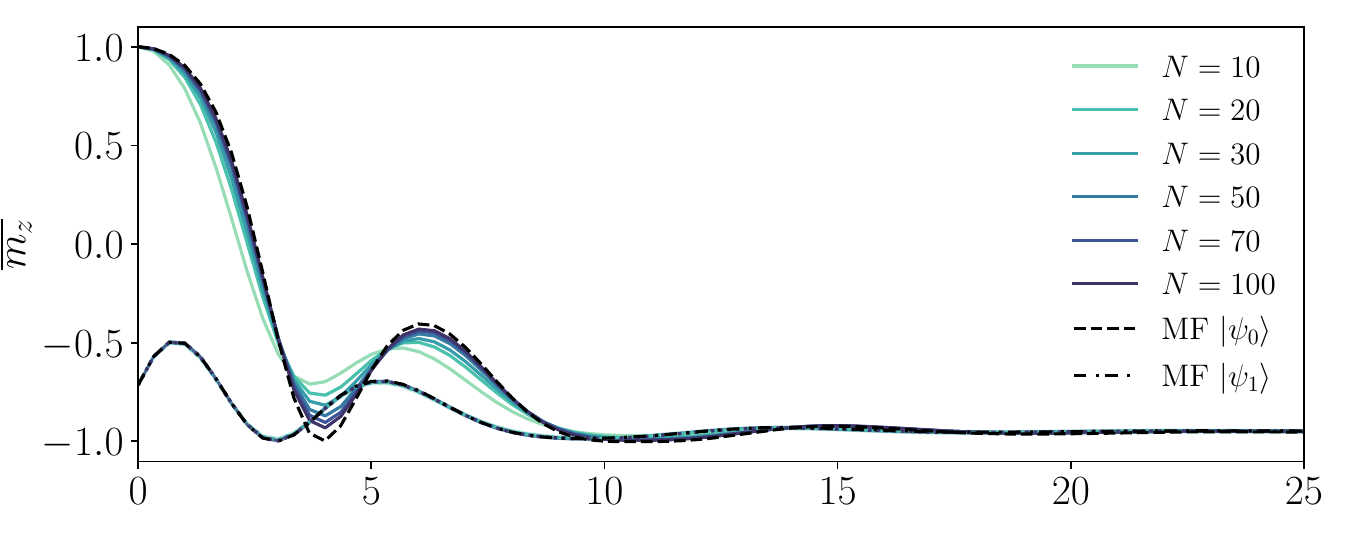} \put(-3,34) {(a)}
    \end{overpic}}
    {\begin{overpic}
    [width=1\linewidth]{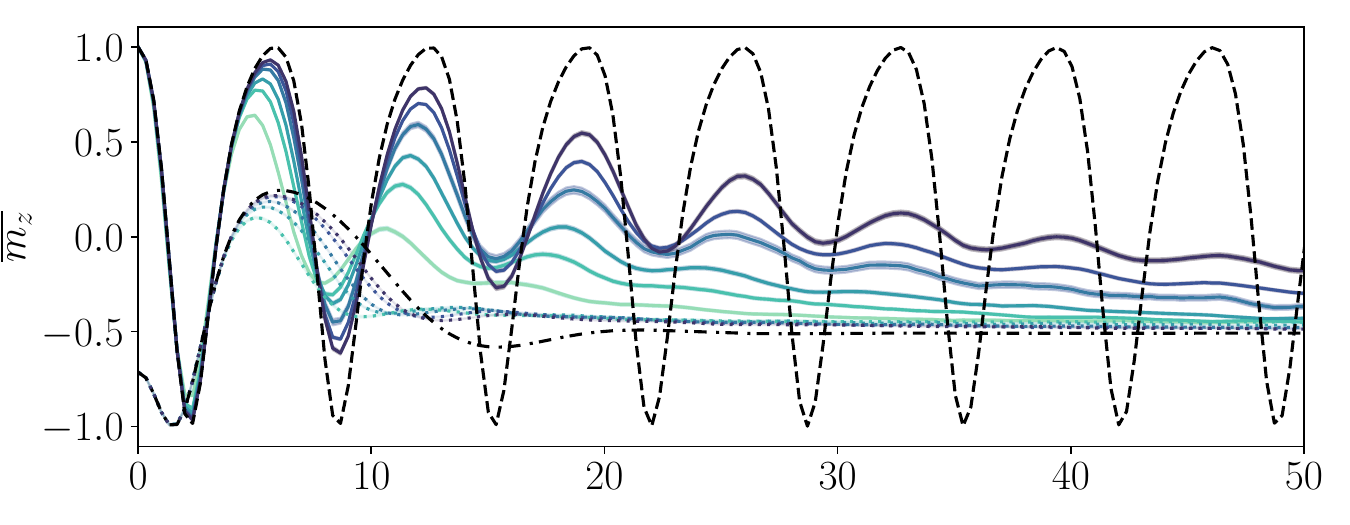}
    \put(-3,34) {(b)}
    \end{overpic}}
    {\begin{overpic}
    [width=1\linewidth]{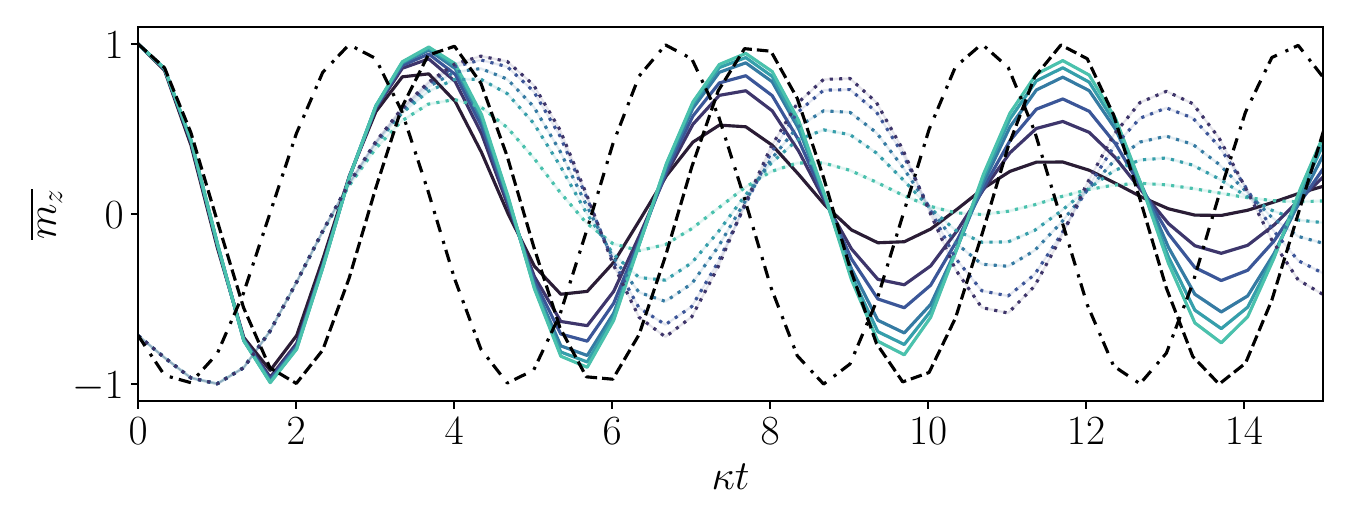}\put(-3,34) {(c)}
    \end{overpic}}
    \caption{$\overline{m_z}$ from initial states $\ket{\psi_0}$ (solid lines) and $\ket{\psi_1}$ (dotted lines) for different $N$ (corresponding to gradient colors) compared to the mean-field for $\ket{\psi_0}$ (black dashed line) and $\ket{\psi_1}$ (black dotdashed line). (a), (b) and (c) are respectively $\Omega/\kappa=0.2,0.55,0.8$. All the plots are obtained from $N_{\mathrm{traj}}=5000$ trajectories and the legend is shared.}.  
    \label{figAnna_6}
\end{figure}

The analysis of quantum correlations in this region is strongly affected by the bistable nature of the model. Therefore, it is convenient to start by carefully analyzing the finite-size dynamics of the magnetization. In the following plots, the bistable region will be signaled by a light yellow shadowed region.
 
Fig.~\ref{figAnna_6} shows the time-evolution of the magnetization starting from two initial states: $\ket{\psi_0}=\ket{m_y=0, m_z=1, m_q=0}=\ket{0}\otimes\ket{\uparrow,\uparrow,...,\uparrow}$ and $\ket{\psi_1}=\ket{m_y=0.7, m_z=-0.71, m_q=-1}$, for which the mean-field phase diagram predicts a phase transition respectively at $\Omega_C^{(0)}/\kappa=0.46$ and $\Omega_C^{(1)}/\kappa=0.57$. The plot presents their dynamics at finite sizes compared to the mean-field one. The finite-$N$ magnetization's evolution in the stationary and time-crystal phases shows the same behavior starting from the two initial states and matches the predictions of the mean field. See [panels (a) and (c)]. [Panel (b)] shows instead the dynamics in the bistable region for $\Omega/\kappa=0.55$, a value between the critical points of the two states. The finite size magnetization has an interesting behavior in this case: $\ket{\psi_0}$ yields a magnetization with oscillations in a transient time, and then a decay to a steady-state value; $\ket{\psi_1}$ shows a transient stationary behavior which then leads to the same steady-state value as $\ket{\psi_0}$.
Furthermore, the duration of the transient increases as the size is increased, and in this time interval the finite-size magnetizations match their respective mean-field predictions: oscillatory for $\ket{\psi_0}$ and stationary for $\ket{\psi_1}$. 

This indicates the non-commutativity of the thermodynamic limit and the infinite-time limit 
\begin{equation}
\lim_{t\to\infty}\lim_{N\to\infty}\overline{m_z(t)}\not =\lim_{N\to\infty}\lim_{t\to\infty}\overline{m_z(t)}.
\end{equation} 
Taking the thermodynamic limit first makes the transient dominate the dynamics and yields a different behavior at any time for the two initial states. On the contrary, taking the infinite time limit first makes the unique steady state dominant (independent of the choice of the initial states). This point is especially important since we are going to discuss properties like the entanglement entropy that should be computed at finite $N$ in the long-time limit.

\begin{figure}
    \centering
    {\begin{overpic}
    [width=.43\linewidth]{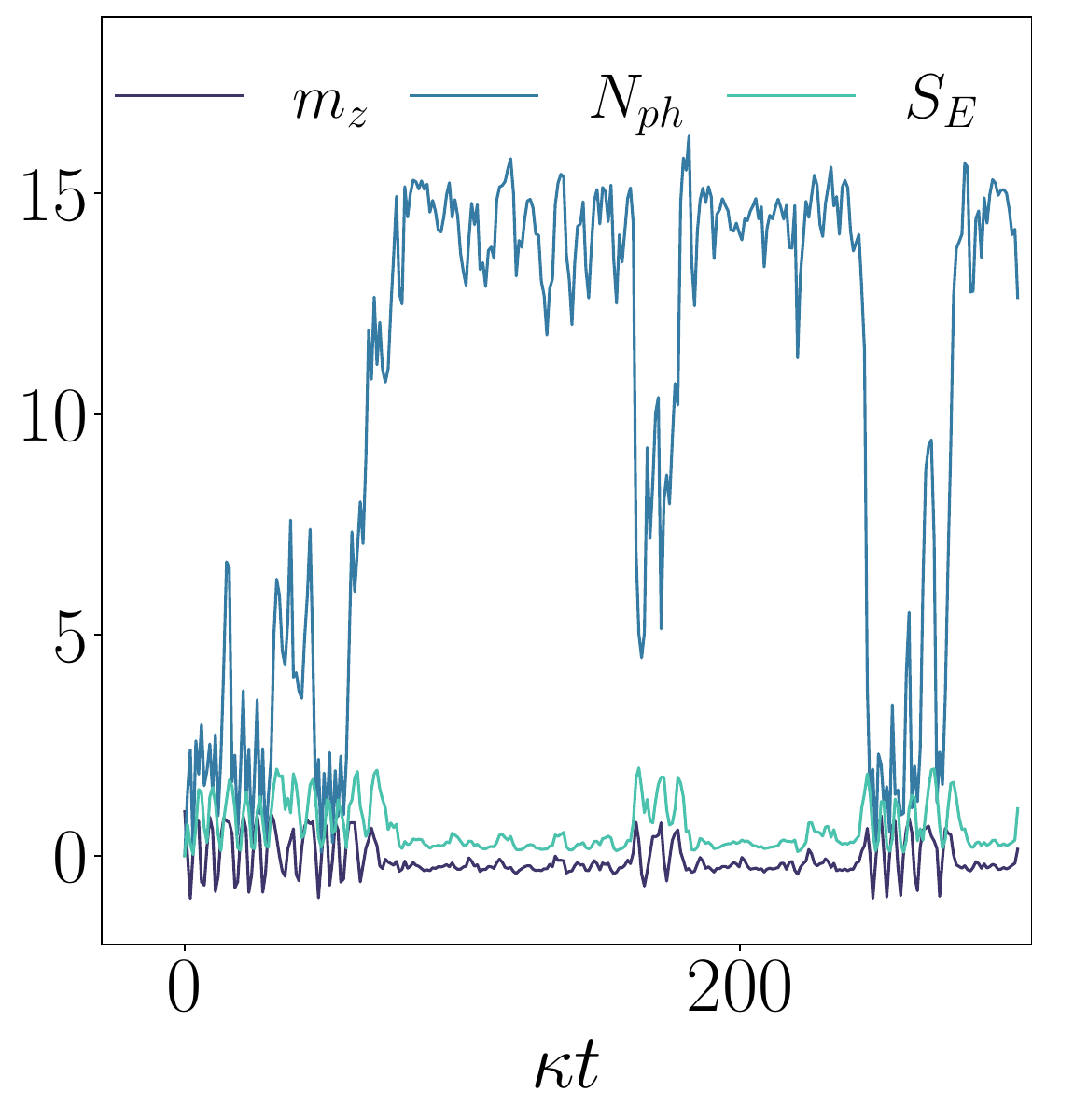} \put(-10,80) {(a)}
    \end{overpic}}
    {\begin{overpic}
    [width=.45\linewidth]{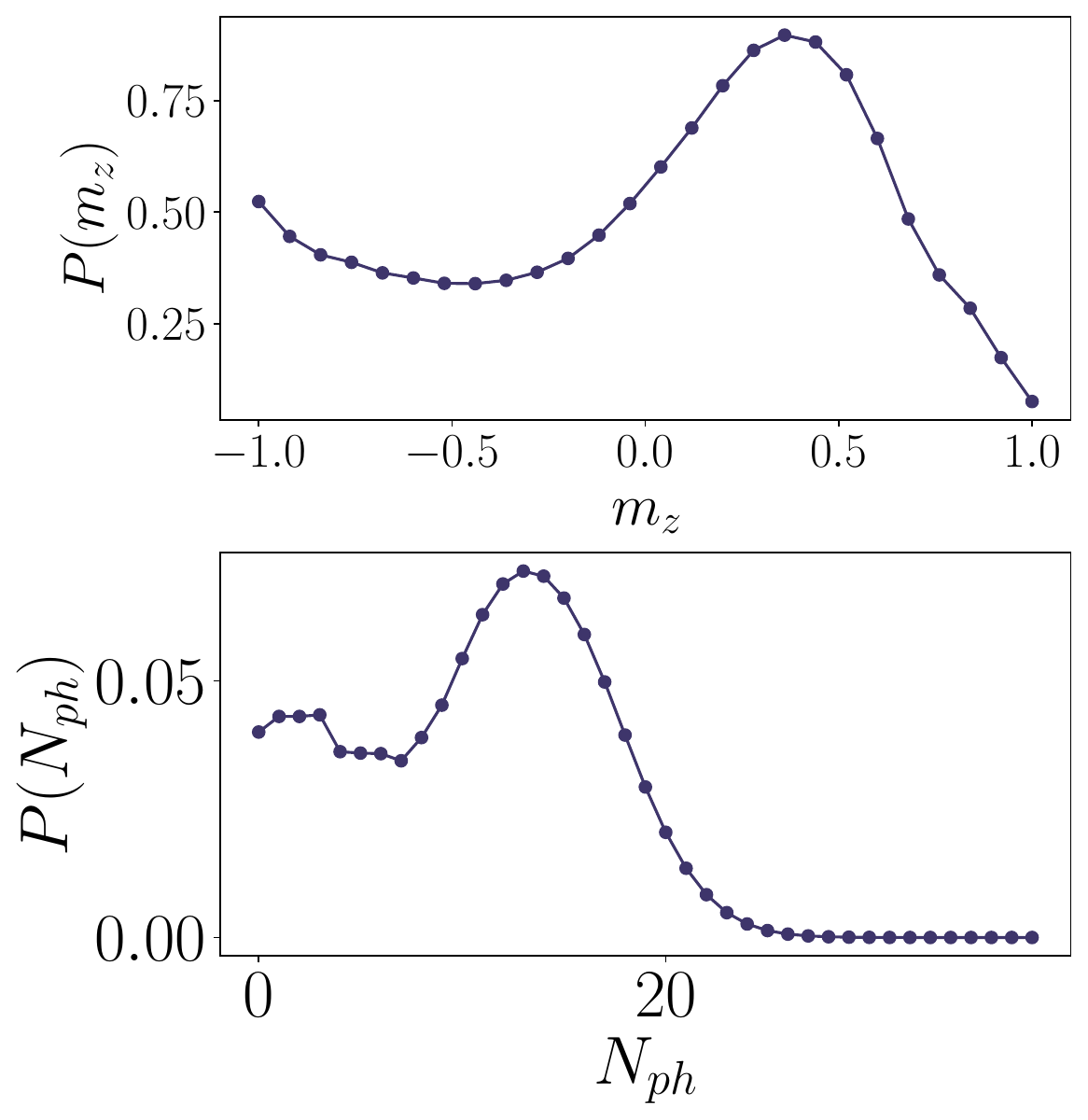}
    \put(20,90) {(b)} \put(20,40) {(c)}
    \end{overpic}}
    \caption{Plots for $N=25$, and  $\Omega/\kappa=0.62$. (a) Single-trajectory values for the average magnetization, number of photons, and entanglement entropy in the bistable phase. (b) and (c) show the bimodality in the probability distribution of the magnetization and the number of photons in the steady-state.}.  
    \label{figAnna_plusplus}
\end{figure}

Bistability is not only witnessed in averaged observables but also at a single-trajectory level. Along one trajectory we notice the switching between two different behaviors: an oscillating highly-entangled one with a low average number of photons $N_\text{ph}=\expval{\hat a^\dag \hat a}$, and a stationary one with a high number of photons which is separable between spins and photons, see Fig.~\ref{figAnna_plusplus}. In the bistable phase, for $N$ finite, the lifetime of the states between which we observe a switching is finite and comparable. In the stationary and time-crystal phases, one of the two behaviors dominates while the other gets suppressed, this yields a commutativity of the thermodynamic and infinite-time limit outside the bistable region. Bistability can also be noticed as bimodality in the steady-state probability distribution of the observables $m_z$ and $N_\text{ph}$~\cite{Schiro}, see panels (b) and (c) of Fig.~\ref{figAnna_plusplus}.

An important consequence of the non-commutativity of the two limits is that the phase diagram of the model is twofold: it can be obtained with the limit ordering $\lim_{t\to\infty}\lim_{N\to\infty}$, corresponding to a `dynamical' phase diagram that only studies the transient time and leads to the magnetization mean-field phase diagram in Fig.~\ref{figAnna_2}; it can be obtained with the opposite ordering $\lim_{N\to\infty}\lim_{t\to\infty}$, corresponding to the phase diagram of the steady-state, independent on the choice of the initial state since the steady state is unique for finite sizes.

To obtain the \textit{dynamical} and \textit{steady-state} phase diagram the following procedures are adopted. The decay time to the steady state $\tau_m$ is obtained from the time evolution of the magnetization, see Appendix~\ref{app:sat}. Choosing $\mathcal T=3\tau_m$ as the boundary between transient and steady state, one can simply define the order parameter for the \textit{steady-state} phase diagram as:
\begin{equation}\label{eqAnna:op_ss}
    \tilde m_z^\text{SS} = \lim_{t_f\to\infty}\frac{1}{(t_f-\mathcal T)}\int_{\mathcal T}^{t_f} dt \, \overline m_z (t).
\end{equation}
The order parameter for the \textit{dynamical} phase diagram, can be instead defined as
\begin{equation}\label{eqAnna:op_dyn}
    \tilde m_z^\text{Dyn} = \frac{1}{\mathcal T}\int_0^\mathcal T dt \, \overline m_z (t).
\end{equation}
The latter definition of order parameter leads to the already presented mean-field phase diagram and therefore will not be presented in this section.\\

\begin{figure}
    \centering
    {\begin{overpic}
    [width=0.9\linewidth]{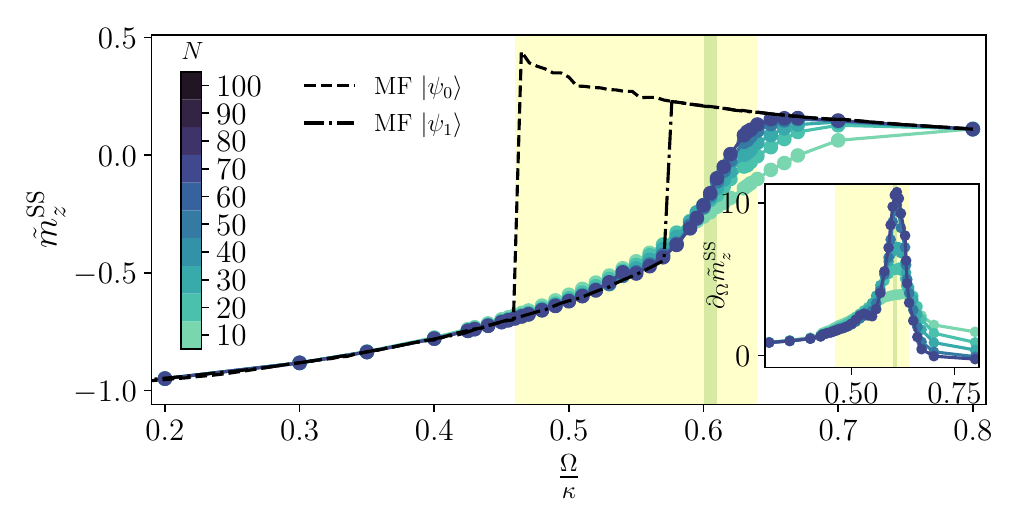} \put(0, 44) {(a)}
    \end{overpic}}
    \caption{Steady-state phase diagram for finite size compared with mean-field phase diagram for $\ket{\psi_0}$ (dashed line) and $\ket{\psi_1}$ (dash-dot line). Derivative of $\tilde m_z^\text{SS}$ in the inset. The green narrow region signals the critical point. The yellow shadowed region  corresponds to the bistable phase}
    \label{figAnna_7}
\end{figure}

The \textit{steady-state} phase diagram and its comparison with the mean-field (or \textit{dynamical} for $N\to\infty$) phase transitions for $\ket{\psi_0}$ and $\ket{\psi_1}$ are presented in Fig.~\ref{figAnna_7}. The steady-state magnetization has a behavior that does not match any of the mean-field first-order transitions. The dark-shadowed green line signals the value of $\Omega_C^\text{SS}/\kappa\sim0.61$ in which the $\Omega$-derivative of $\tilde m_z^\text{SS}$ has a maximum and separates two regions with positive and negative magnetizations. We identify this value as critical point for the magnetization phase transition. In all plots (Figs. 9,11-13)  the bistable region  is indicated  with a yellow shadowed
region.

\begin{figure}
    \centering
    {\begin{overpic}
    [width=1\linewidth]{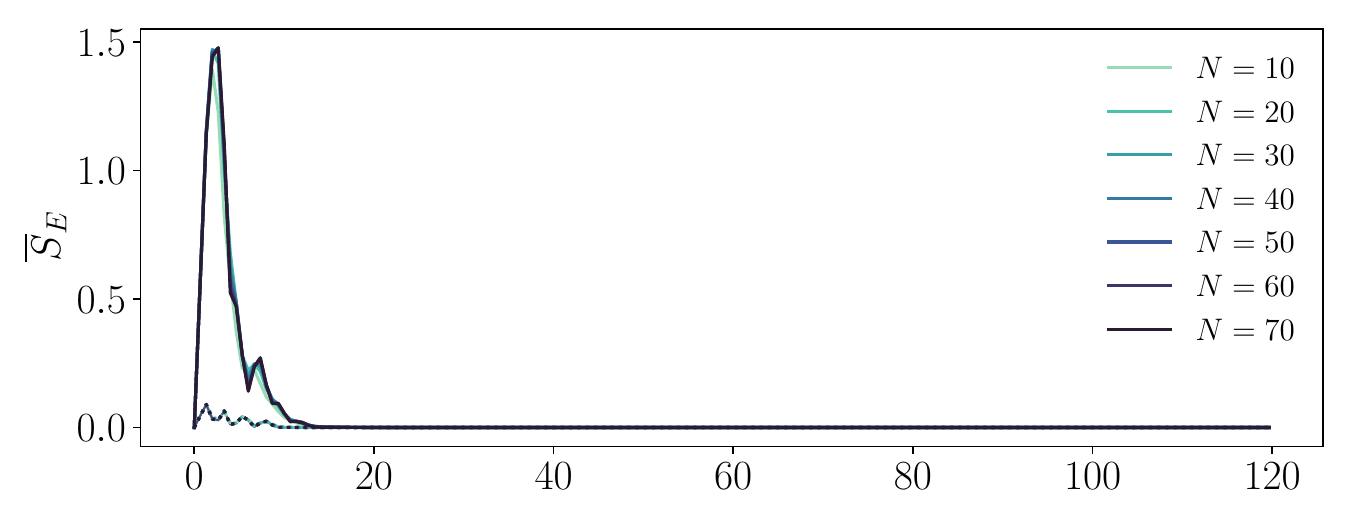} \put(0,34) {(a)}
    \end{overpic}}
    {\begin{overpic}
    [width=1\linewidth]{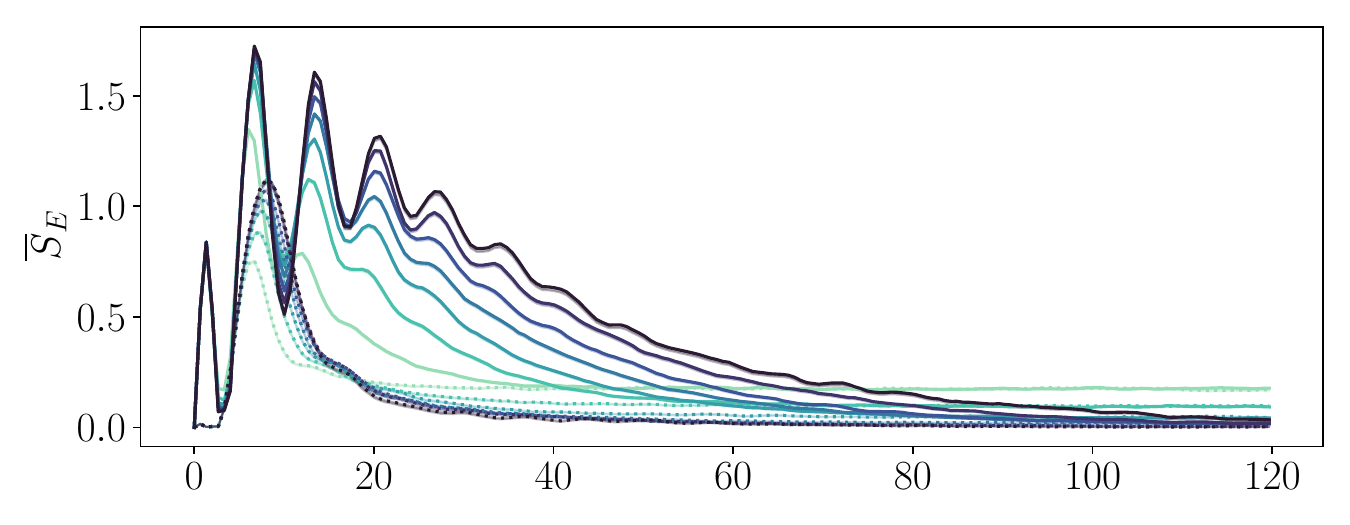}
    \put(0,34) {(b)}
    \end{overpic}}
    \begin{overpic}
    [width=1\linewidth]{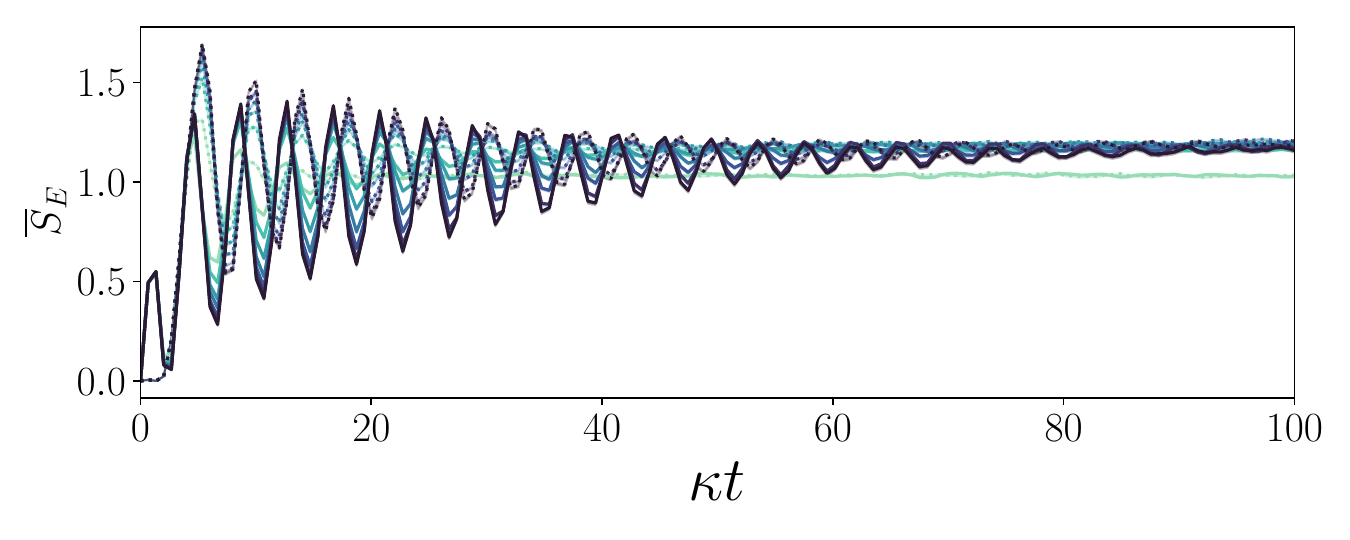}
    \put(0,34) {(c)}
    \end{overpic}
    \caption{$\overline S_E$ from initial states $\ket{\psi_0}$ (solid lines) and $\ket{\psi_1}$ (dotted lines) for different $N$. (a), (b), and (c) are respectively for $\Omega/\kappa=0.2,0.55,0.8$. All the plots share the legend and are obtained from $N_{\mathrm{traj}}=5000$ trajectories.}  
    \label{figAnna_8}
\end{figure}                                                  

The same procedure is applied again to analyze properly the entanglement entropy. The time-evolution of the entanglement entropy in Fig.~\ref{figAnna_8}, obtained from $\ket{\psi_0}$ and $\ket{\psi_1}$, shows once again the non-commutativity of the thermodynamic limit and infinite time limit in the bistable region. 

Also in this case, a \textit{steady-state} and a \textit{dynamical} order parameter can be defined. Following closely the procedure already used for the magnetization, we can identify a $\tau_S$ saturation time from the time-evolution of the entanglement entropy, set $\mathcal T=3\tau_S$ and define $\tilde S_E^\text{SS}$ with a time-average over $[\mathcal T,\infty]$ and $\tilde S_E^\text{Dyn}$ through a time-average over the window $[0,\mathcal T]$.

\begin{figure*}
    \centering
    {\begin{overpic}
    [width=0.47\linewidth]{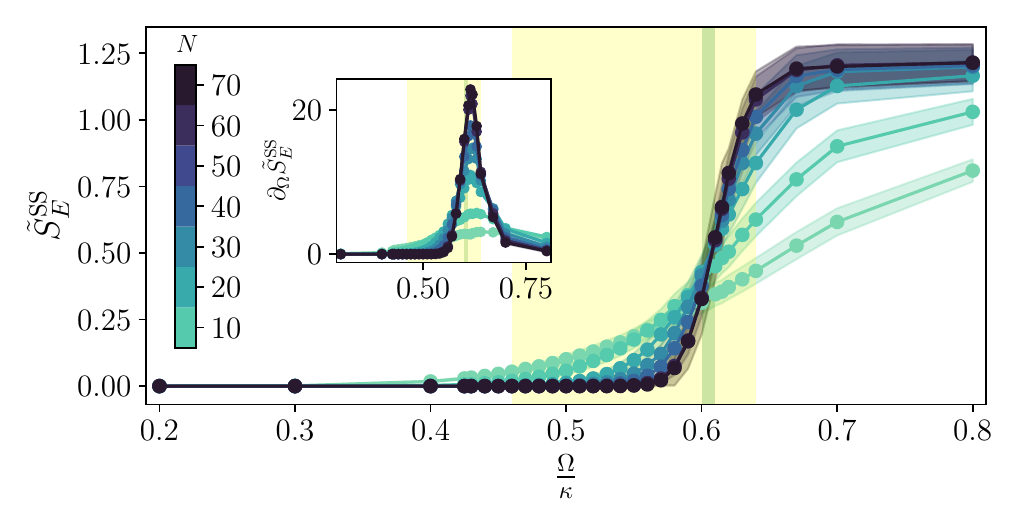} \put(0,44) {(a)}
    \end{overpic}}
    {\begin{overpic}
    [width=0.245\textwidth]{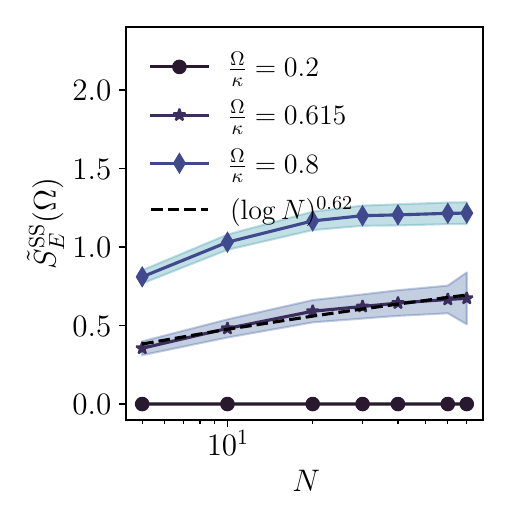}   \put(0,85) {(b)}
    \end{overpic}}
    {\begin{overpic}
    [width=0.245\textwidth]{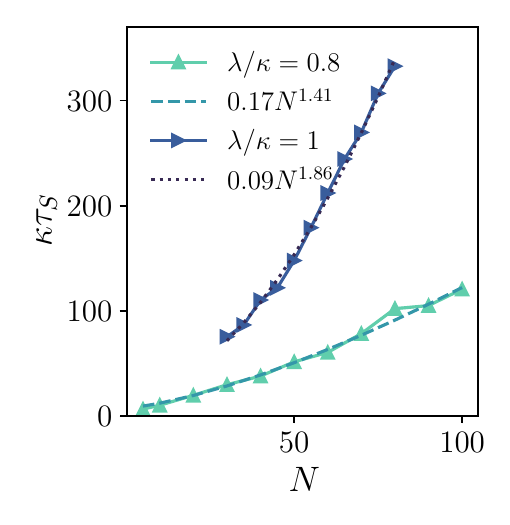}\put(0,85) {(c)}
    \end{overpic}}
    \caption{(a) $\tilde S_E^\text{SS}$ for different size and its derivative in the inset. The green narrow region signals the critical point. (b) Scaling of $\Tilde S_E$ with $N$ in the steady-state phase diagram. (c) Scaling of the entanglement entropy saturation time with system size $N$ at criticality. The lines correspond to two different values of $\lambda/\kappa$ both affected by the bistability. The yellow region corresponds to the bistable phase 
    } 
    \label{figAnna_9}
\end{figure*}   

The unique \textit{steady-state} phase diagram in Fig.~\ref{figAnna_9}(a) shows an entanglement transition between a non-entangled region in which $\tilde S_E^{\rm{SS}}\propto N^0$ and an entangled region with $\tilde S_E^{\rm{SS}}\propto N^0$ (scalings shown in [panel (b)]). The green shadowed line, that separates the two phases, indicates the value $\Omega_C^{\rm{SS}}\sim0.61$ at which the derivative has a maximum. This value precisely coincides with what we found in the magnetization \textit{steady-state} phase diagram. Once again we have found in the driven-dissipative Tavis Cummings-Model an entanglement transitions occurring together with a magnetization transition.
However, an important difference arises from the previous entanglement phase transition found at $\lambda/\kappa=0.2$. The saturation time at criticality scales in this case superlinearly, without mitigating the post-selection barrier. Fig.~\ref{figAnna_9}(c) shows the saturation time at criticality for $\lambda/\kappa=0.8$ and deeper in the bistable region at $\lambda/\kappa=1$: in both cases, the entanglement presents a slow saturation worsened by the increase of $\lambda/\kappa$.

The \textit{dynamical} computation of $\tilde S_E^\text{Dyn}$ shows for both initial states a jump in the entanglement entropy around a critical point that matches exactly the critical point for the magnetization transition in the mean-field (\textit{dynamical}) phase diagrams. We conclude that also in this different type of phase diagram, we have an entanglement transition occurring along with the magnetization one. The scalings of the entanglement entropy in the entangled and non-entangled region of the \textit{dynamical} phase transition are shared with the \textit{steady-state} one, as outside the bistable region the commutativity of the thermodynamic limit and infinite time limit holds.

\begin{figure}
    {\begin{overpic}
    [width=0.454\linewidth]{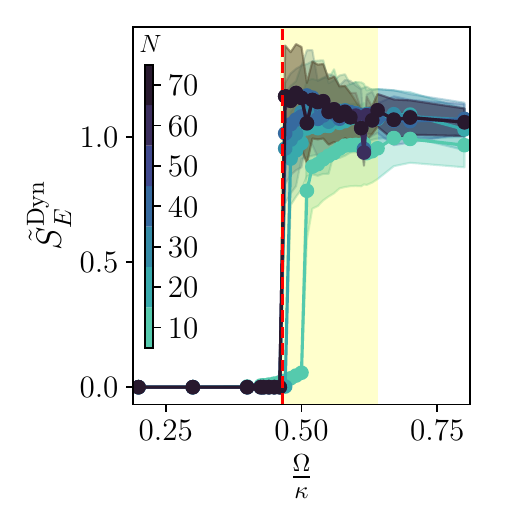} \put (0,85) {(a)}
    \end{overpic}}
    {\begin{overpic}
    [width=0.454\linewidth]{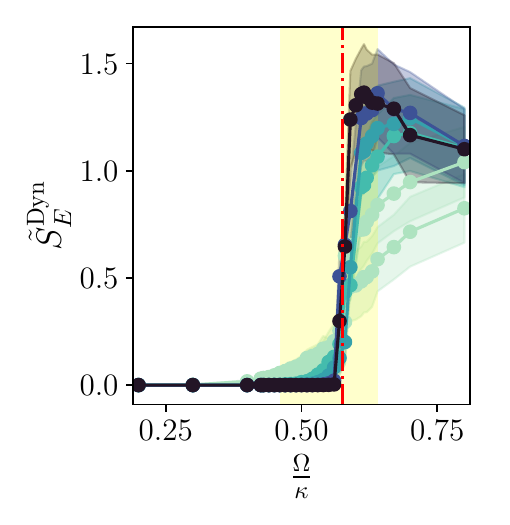} \put(1,85) {(b)}
    \end{overpic}}
    \caption{(a) $\tilde S_E^\text{Dyn}$ calculated from $\ket{\psi_0}$ and critical value as red dashed line. (b) $\tilde S_E^\text{Dyn}$ calculated from $\ket{\psi_1}$ and critical value as red dot-dash line. The yellow shadowed region corresponds to the bistable phase and the legend is shared.
    } 
    \label{figAnna_10}
\end{figure}

\begin{figure}
    \centering
    \includegraphics[scale=0.7]{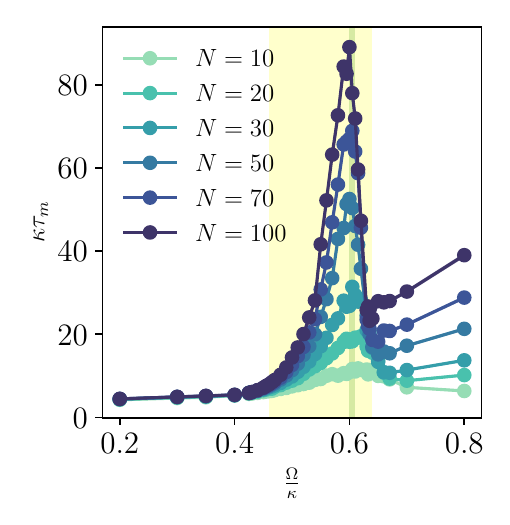}
    \caption{Saturation time of the magnetization for different sizes. The plot is obtained for $\ket{\psi_0}$, but the same holds from $\ket{\psi_1}$. The yellow part corresponds to the bistable phase.} 
    \label{figAnna_11}
\end{figure}  

The features arising from the bistability of the model can be summed up in a single plot. Fig.~\ref{figAnna_11} shows the behavior of the magnetization's saturation time with respect to $\Omega/\kappa$. Before the bistable region (yellow shadowed region) the saturation time does not have size-dependence and this plateau corresponds to a minimum, this identifies the stationary phase (in the mean-field \textit{dynamical} sense). In the bistable region, the saturation time starts to have a size dependence and peaks around $\Omega_C^\text{SS}/\kappa\sim0.61$, which can be identified as the critical point of the \textit{steady-state} phase diagram and decreases until the end of the bistable region. A cusp appears at the boundary between the bistable and time-crystal regions (in the mean-field \textit{dynamical} sense) and then starts to increase again. Thus, the cusp defines where the bistable region ends.
This plot allows us to identify altogether the three regions of the \textit{dynamical} phase diagram, along with the boundary between the stationary phase and time-crystal phase of the \textit{steady-state} phase diagram.

\subsection{Driven spins with collective absorption and
emission}\label{res_finiteTSpin}

In our finite-$N$ simulations, we consider $N_\text{traj} = 1024$ trajectories. We initialize the system in the state $\ket{\psi(0)} = \ket{\uparrow\uparrow \dots \uparrow}$. We consider temperatures in the range $T\in [0, 2\kappa]$, for three values of $\Omega$ representative of the whole phase diagram: $\Omega = \kappa/2$ (magnetized phase), $\Omega = \kappa$ (critical point), and $\Omega = 3\kappa/2$ (TC phase). \textit{En passant}, we note that the finite temperature is responsible for finite-size effects in the behavior of the order parameter of the symmetry-breaking transition, $\tilde{m}_z$, which disappear for large $N$.

\begin{figure*}[t]
    \centering
    \includegraphics[width = \textwidth]{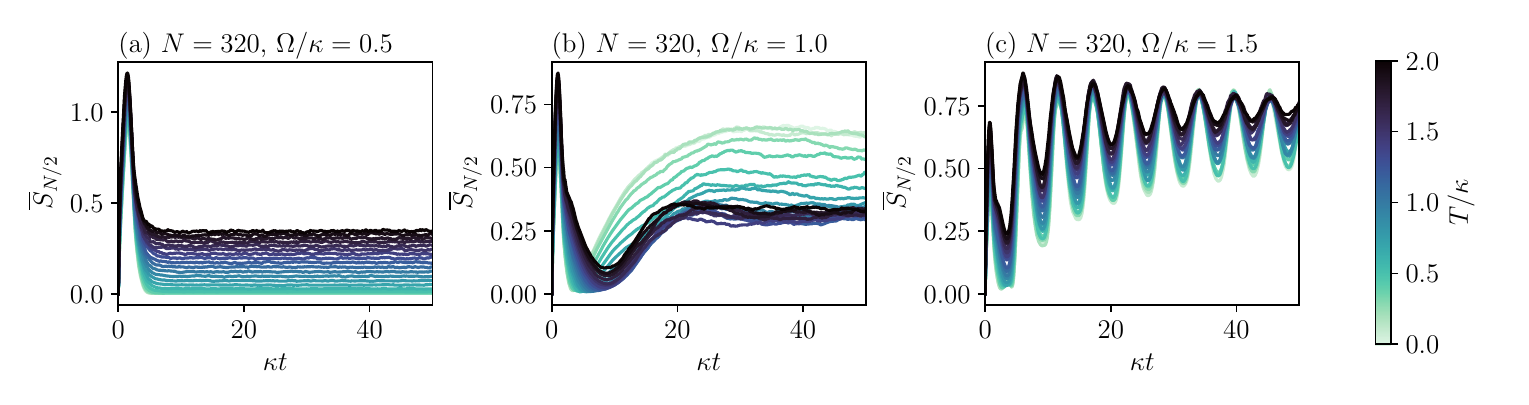}
    \caption{Entanglement dynamics for $N = 320$ in the finite-temperature superradiant model, for different temperatures (see colorbar).}
    \label{fig:btc-finite-t-entanglement-dynamics}
\end{figure*}

First of all, we study the effect of raising the temperature for a fixed system size. In Fig.~\ref{fig:btc-finite-t-entanglement-dynamics}, we plot $\overline{S}_{N/2}$ as a function of time for $N = 320$. For a single trajectory with index $\gamma$, the half-system entanglement entropy is defined as 
\begin{equation}
    S_{N/2}^{(\gamma)} = -\Tr \hat\rho_{N/2}^{(\gamma)} \log \hat\rho_{N/2}^{(\gamma)},
\end{equation}
where $\hat\rho^{(\gamma)}_{N/2}$ is the reduced density matrix obtained from the pure-state density matrix $\hat{\rho}_{\gamma} = \ket{\psi_\gamma}\bra{\psi_\gamma}$ by tracing out half of the system, irrespective of the single-particle indices thanks to permutational invariance~\cite{latorre2005entanglementLMG}.
In [panel (a)], we show the behavior of $\overline{S}_{N/2}$ in the magnetized phase. We see that the entanglement entropy quickly reaches a plateau that corresponds to the stationary state. The value at the plateau depends on $T$: jumps mediated by $\hat S_+$ push the state vector up the Dicke ladder, towards states with larger entanglement entropy compared to the bottom of the ladder. As a result, the value at the plateau grows with $T$. In [panel (b)], we plot the dynamics of entanglement at the MF critical point. Also in this case, the entanglement entropy quickly saturates to a plateau for all values of $T$. However, this time increasing the temperature from $T = 0$ makes the stationary value of $\overline{S}_{N/2}$ decrease. By further increasing $T$ (above $T \approx 0.5\kappa$), the stationary value no longer changes. In [panel (c)], we show the entanglement dynamics in the TC phase. Here, we see that the effect of raising the temperature is to decrease the amplitude of the oscillations of $\overline{S}_{N/2}$, which are related to the oscillations of $m_z$ in the time-crystal phase.

\begin{figure*}[t]
    \centering
    \includegraphics[width = \textwidth]{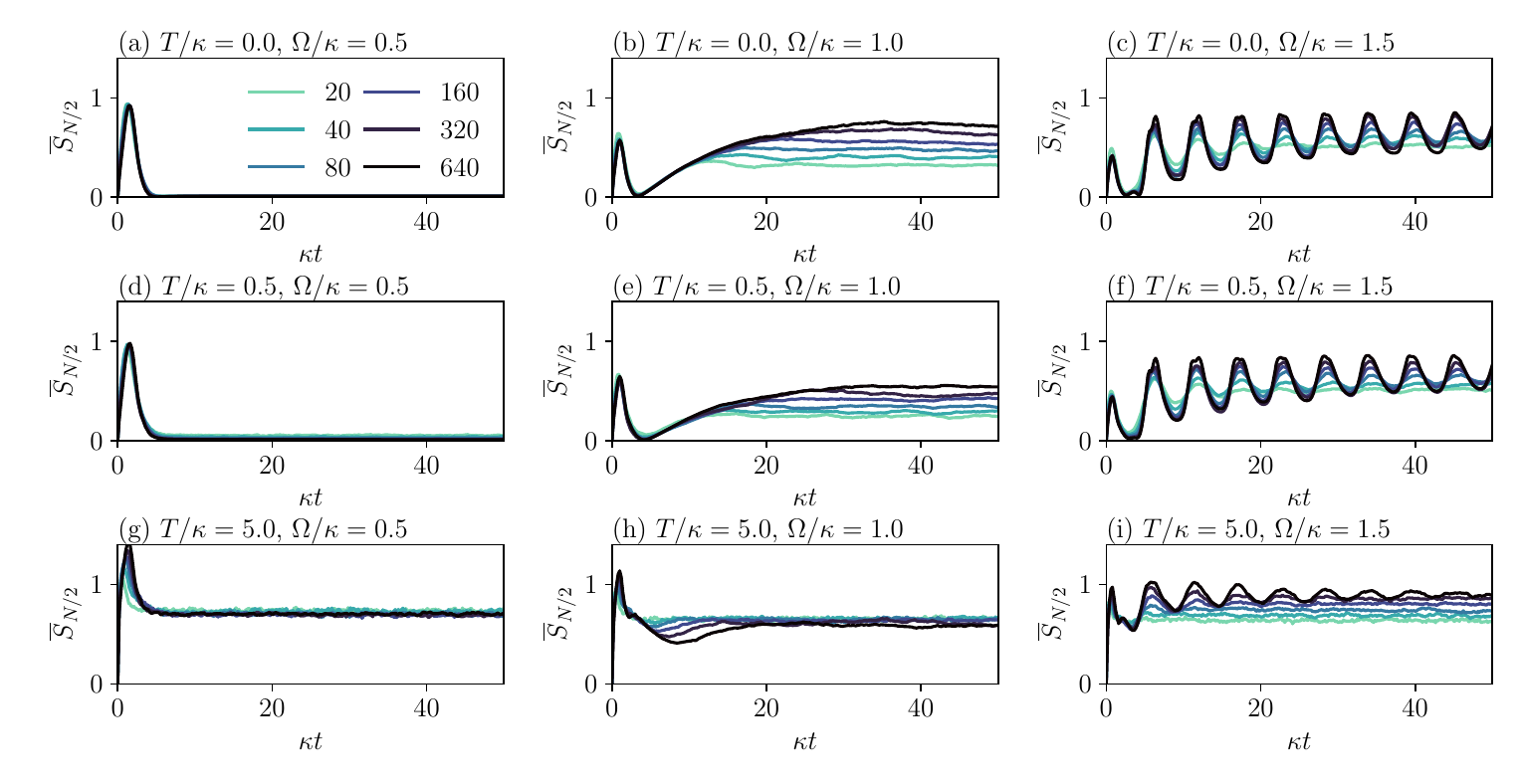}
\caption{Entanglement dynamics in the finite-$T$ superradiant model, for different temperatures, frequencies, and sizes [sizes are shown in the legend of panel (a)].}
    \label{fig:btc-finite-t-entanglement-dynamics-size}
\end{figure*}

Second, we study the entanglement entropy for several sizes $N$, and for some temperatures $T$, to see how the finite temperature affects its scaling properties. We report our results in Fig.~\ref{fig:btc-finite-t-entanglement-dynamics-size}. First of all, let us recall the results of the zero-$T$ model (top row) \cite{Passarelli}. In all phases, $\overline{S}_{N/2}$ reaches a stationary value within a time frame that scales differently with $N$, depending on $\Omega / \kappa$. In the magnetized phase (left), the entanglement entropy obeys an area law, and the saturation time is size-independent. At the transition (center), both the asymptotic entanglement entropy $\tilde{S}_{N/2}$ and the saturation time scale as $\log N$. In the time-crystal phase (right), the entanglement entropy is sublogarithmic and the saturation time grows as $\log N$ (with oscillations superimposed on the saturation behavior). Therefore, the analysis of the quantum trajectories unveils a measurement-induced entanglement transition whose critical point coincides with the one of the MF transition~\cite{Passarelli}.

\begin{figure*}[t]
    \centering
    \includegraphics[width = \textwidth]{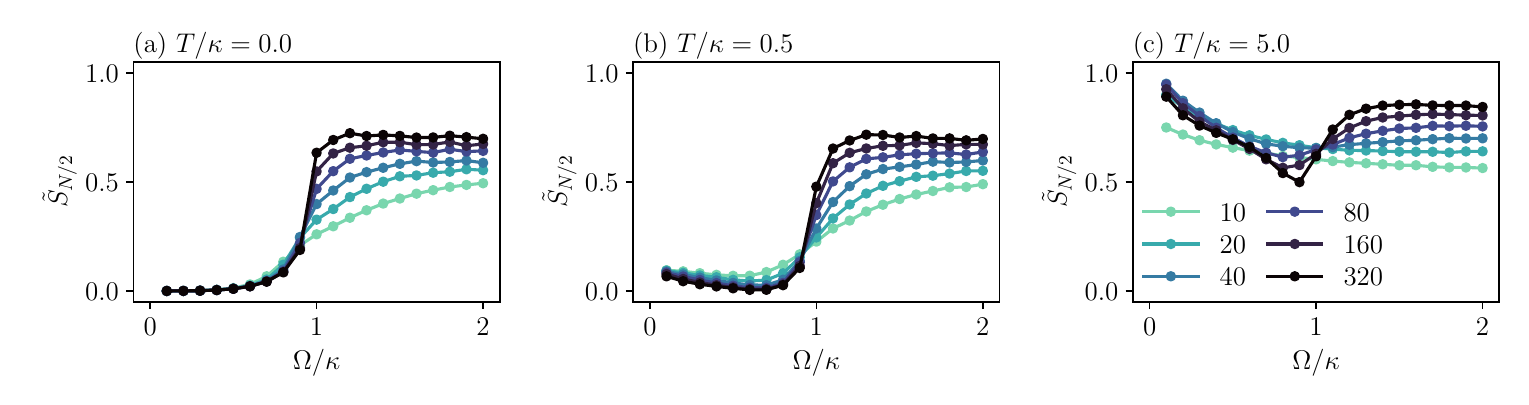}
    \caption{Entanglement entropy of the finite-$T$ superradiant model at saturation as a function of $\Omega$, for different temperatures and sizes [see legend of panel~(c)].}
    \label{fig:btc-finite-t-entanglement-scaling}
\end{figure*}

Some of these features are found also at finite temperatures, while others do change. In the other two rows of Fig.~\ref{fig:btc-finite-t-entanglement-dynamics-size}, we consider $T = \kappa / 2$ and $T = 5 \kappa$. In the magnetic phase (left column), the effect of a finite $T$ is to increase the stationary value of the entanglement entropy as discussed before; the system remains in an area law phase and the saturation time remains independent of $N$. At the critical point (center column), we see that by increasing the temperature the $\log N$ dependence of the asymptotic entanglement entropy $\tilde{S}_{N/2}$ is lost for sufficiently high temperatures and all curves saturate, in a time scale proportional to $\log N$ (see later), to a size-independent (area) plateau. Finally, in the TC phase (right column), the finite temperature suppresses the time-crystal oscillations of $\overline{S}_{N/2}$ and the scaling of the asymptotic entanglement entropy becomes $\log N$. This is more clearly seen in Fig.~\ref{fig:btc-finite-t-entanglement-scaling}, where we report the stationary $\tilde{S}_{N/2}$ as a function of $\Omega$, for different system sizes and at different temperatures. We can clearly see that, in the magnetized phase, the scaling with $N$ is unchanged compared to $T = 0$. Instead, the logarithmic scaling of $\tilde{S}_{N/2}$ with $N$ at criticality is fragile and disappears by increasing $T$: for instance, at $T = 5\kappa$, all curves cross close to $\Omega = \kappa$, where there is no scaling with $N$ within margins of error. Apart from this, also at finite temperatures, we observe a measurement-induced entanglement transition, separating an area-law phase ($\Omega < \kappa$) from a sub-volume phase ($\Omega > \kappa$). 

\begin{figure}[t]
    \centering
    \includegraphics[width = \columnwidth]{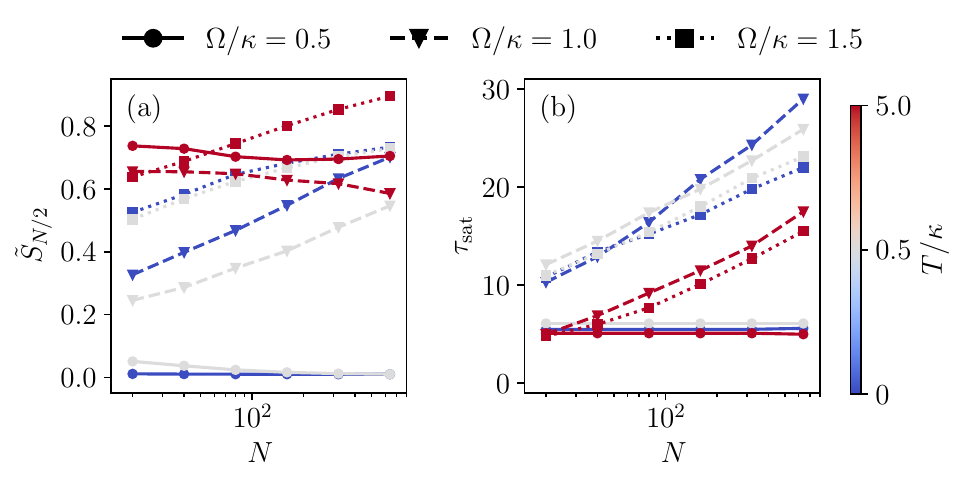}
    \caption{(a) Entanglement entropy at saturation and (b) saturation time as a function of $N$, for different temperatures and $\Omega$ (finite-$T$ superradiant model).}
    \label{fig:btc-finite-t-entanglement-scaling-time}
\end{figure}

We summarize the scaling of $\tilde{S}_{N/2}$ and its saturation time $\tau_\text{sat}$ with the system size in Fig.~\ref{fig:btc-finite-t-entanglement-scaling-time}, for some values of $T$ and $\Omega$.

\subsection{Driven-dissipative Bose-Hubbard dimer}\label{sec:result-bose-hubbard}

Finally, in this section, we discuss our results on the Bose-Hubbard dimer in the two different regimes introduced in Sec.~\ref{sec:bhd}, namely 1) the regime $J>0$ with a second-order phase transition and 2) the regime $J<0$ with a first-order transition. We will study the dynamics of both linear observables (i.\,e., the usual order parameters for the dissipative phase transitions on the level of the density matrix) and the entanglement entropy. 

\begin{figure*}
    \centering
    \includegraphics[width=\linewidth]{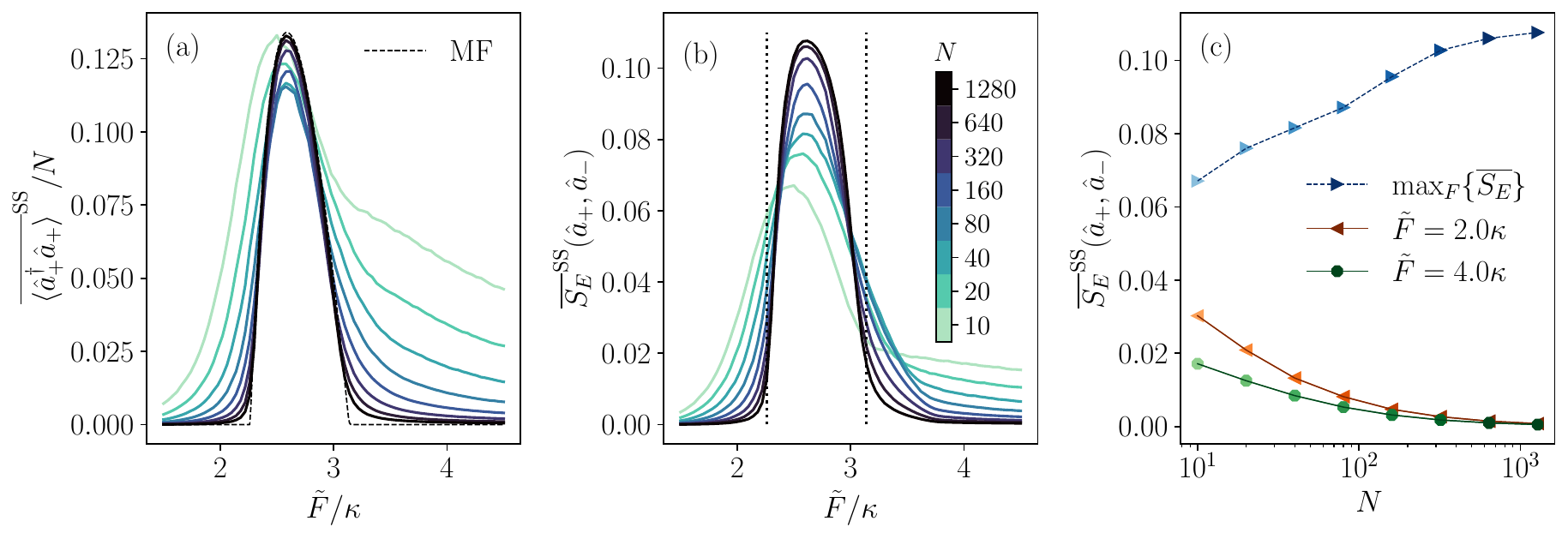}
    \caption{Steady-state behavior of the Bose-Hubbard dimer in the regime with negative drive detuning with respect to the driven mode (the higher one in energy), for different values of the scaling parameter $N$ [legend shared between (a) and (b)]. (a) Normalized population in the bonding mode $\overline{\langle\hat{a}_{+}^\dagger\hat{a}_{+}\rangle}/N$ as a function of the normalized drive amplitude $\tilde{F}$ with the mean-field solution (dashed line) superimposed. (b) Trajectory-averaged entanglement entropy between the two normal modes ($\hat{a}_{+}$ and $\hat{a}_{-}$) of the steady-state as a function of $\Tilde{F}$. The vertical dotted lines mark the critical points predicted by the mean-field theory, at $\tilde{F}=2.26\kappa$ and $\tilde{F}=3.14\kappa$ respectively. (c) Scaling of the maximum steady-state entanglement entropy and for two other values of $\Tilde{F}$ (see legend) versus the scaling parameter $N$  in linear-log scale. Parameters: $J=2.5\kappa$, $\Delta=-1.5\kappa$, $\Tilde{U}=2\kappa$.}
    \label{fig:bhd-ssb-ent}
\end{figure*}

\begin{figure*}
    \centering    \includegraphics[width=\linewidth]{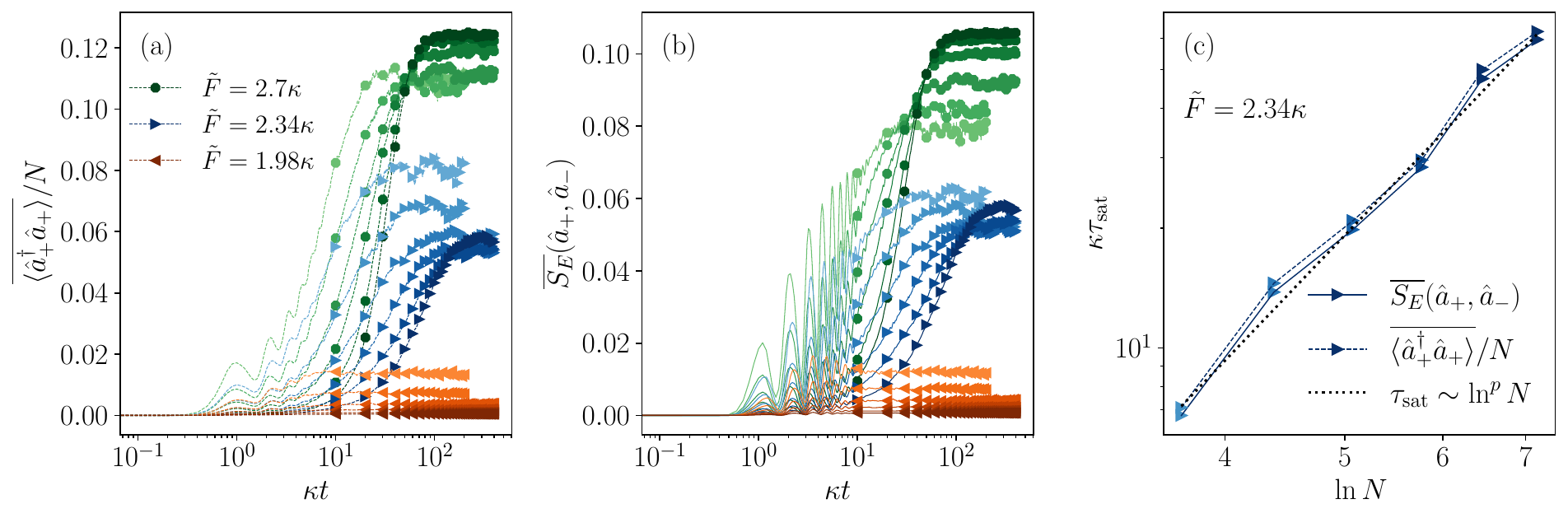}
    \caption{Dynamical properties of the Bose-Hubbard dimer in the regime with negative drive detuning with respect to the driven mode (the higher one in energy), for different values of the scaling parameter $N$ (increasing exponentially from light to darker color: $N=40\,,80\,,160\,,320\,,640\,,1280$). (a) Dynamics of the normalized population of the bonding mode $\hat{a}_{+}$ for three different values of $\Tilde{F}$ (see legend) in linear-log scale. (b) Dynamics of the trajectory-averaged entanglement entropy between the two normal modes $\overline{S}_E(\hat{a}_{+},\hat{a}_{-})$ for the same values of $\Tilde{F}$ as considered in (a), in linear-log scale. (c) Saturation time of the entanglement-entropy (solid line) and of the population (dashed line) at $\tilde{F}=2.34\kappa$ as a function of $N$, in log-loglog scale. The entanglement saturation time is fitted to a power law of $\ln N$: $\tau_{\mathrm{sat}}\sim\ln^{p}N$ with $p= 3.2\pm 0.1$.}
    \label{fig:bhd-ssb-sat}
\end{figure*}

\begin{figure*}
    \centering
    \includegraphics[width=\linewidth]{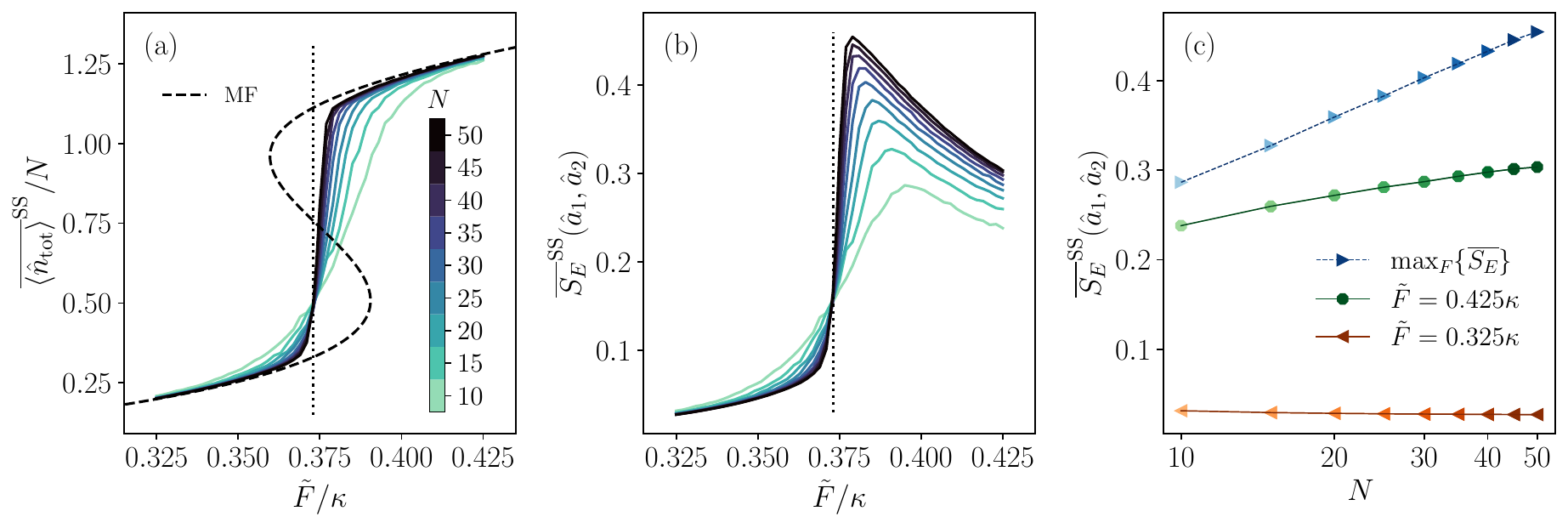}
    \caption{Steady-state behavior of the Bose-Hubbard dimer in the optical bistability regime for different values of the scaling parameter $N$ [legend shared between (a) and (b)]. (a) Normalized total photon population as a function of the normalized drive amplitude $\tilde{F}$ with the mean-field solution (dashed line) superimposed. The vertical dotted line marks the driving amplitude $\tilde{F}=0.373\kappa$, where the curves for different $N$ cross. (b) Trajectory-averaged entanglement entropy between the two spatial modes ($\aaa_1$ and $\aaa_2$) of the steady-state as a function of $\Tilde{F}$. The vertical dotted line is at the same position $\tilde{F}=0.373\kappa$ as in (a), where the different curves for entanglement also cross. (c) Scaling of the maximum steady-state entanglement entropy and for two other values of $\Tilde{F}$ (see legend) versus the scaling parameter $N$  in linear-log scale. Parameters: $J=-2.5\kappa$, $\Delta=-1.4\kappa$, $\Tilde{U}=2\kappa$.}
    \label{fig:bhd-bis-ent}
\end{figure*}

\begin{figure*}
    \centering
    \includegraphics[width=\linewidth]{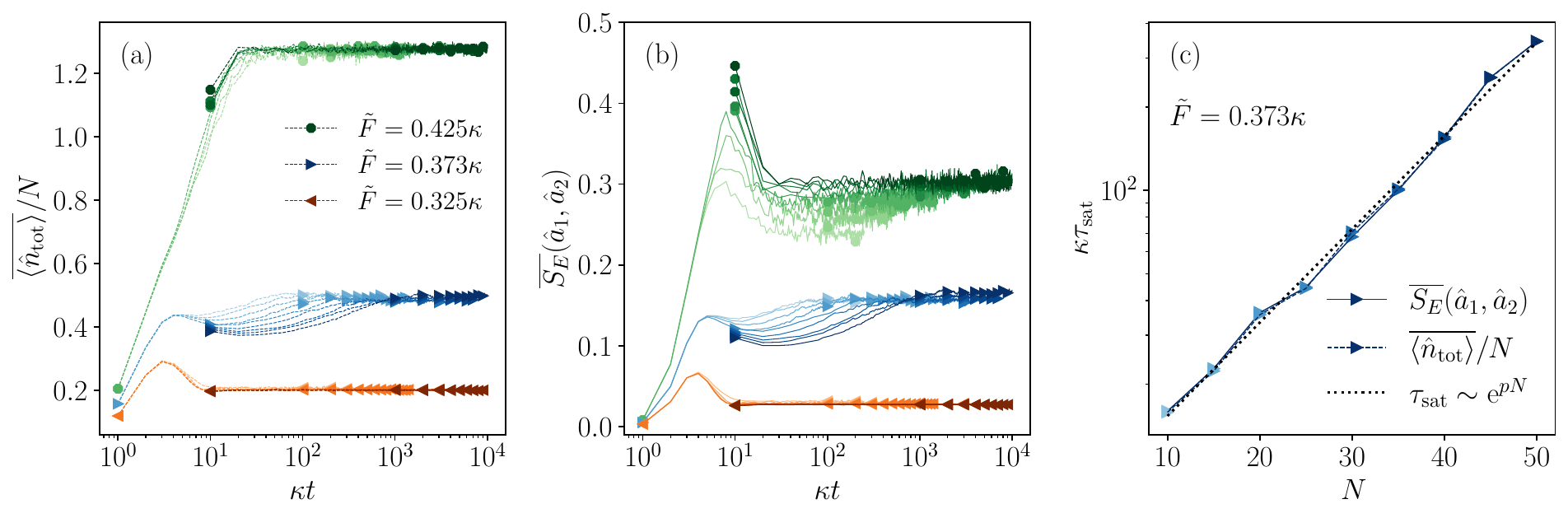}
    \caption{Dynamical properties of the Bose-Hubbard dimer in the optical bistability regime for different values of the scaling parameter $N$ (increasing linearly from light to darker color: $N=10\,,15\,,\cdots\,,50$). (a) Dynamics of the normalized photon population for three different values of $\Tilde{F}$ (see legend) in linear-log scale. (b) Dynamics of the trajectory-averaged entanglement entropy $\overline{S}_E(\aaa_1,\aaa_2)$ for the same values of $\Tilde{F}$ as considered in (a), in linear-log scale. (c) Saturation time of the entanglement-entropy (solid line) and of the population (dashed line) close to the transition as a function of the scaling parameter $N$, in log-linear scale. The entanglement saturation time is fitted to an exponential function $\tau_{\mathrm{sat}}\sim\rme^{pN}$ with $p= 0.078\pm 0.002$.}
    \label{fig:bhd-bis-sat}
\end{figure*}

\subsubsection{$J>0$: Second-order phase transition with fast-saturating entanglement}

We consider the system evolving from a vacuum initial state with parameters $\Delta=-1.5\kappa$, $J=2.5\kappa$ and $\tilde{U}=2\kappa$ (as considered in~\cite{casteelsQuantumEntanglementSpatialsymmetrybreaking2017}). In this regime, the drive has a negative detuning respective to the driven mode $\hat{a}_{-}$ and the system is known to undergo a second-order dissipative phase transition breaking the symmetry between the two sites of the dimer. We study the behavior of the entanglement in the vicinity of this transition using the quantum trajectories within the Gaussian approximation.

Fig.~\ref{fig:bhd-ssb-ent}(a) reproduces the known dissipative phase transition of the average steady state~\cite{garbinSpontaneousSymmetryBreaking2022} witnessed by the order parameter $\overline{\langle\hat{a}_{+}^\dagger\hat{a}_{+}\rangle}/N$ (normalized population in the bonding mode)~\footnote{Note that since the dynamics (such as shown in Fig.~\ref{fig:bhd-ssb-sat}) does not involve limit cycles as in the two other models we considered, our notation for the order parameter here is simply the trajectory average $\overline{\phantom{i}\bullet\phantom{i}}$ without the time average.}. As we approach the thermodynamic limit by increasing the scaling parameter $N$, the steady-state order parameter approaches the mean-field solution. We identify a second-order continuous transition between a symmetric phase with $\overline{\langle\hat{a}_{+}^\dagger\hat{a}_{+}\rangle}/N\xrightarrow[]{N\to\infty} 0$ and a symmetry-broken phase with nonzero population in the bonding mode.

Fig.~\ref{fig:bhd-ssb-ent}(b) shows $\overline{S}_E(\hat{a}_{+},\hat{a}_{-})$, the steady-state trajectory-averaged entanglement between the bonding mode $\hat{a}_{+}$ and the antibonding mode $\hat{a}_{-}$ across this dissipative phase transition. As $N$ increases, we observe similar behavior as the order parameter $\overline{\langle\hat{a}_{+}^\dagger\hat{a}_{+}\rangle}/N$, where the entanglement between the two momentum modes vanishes in the symmetric phase and becomes nonzero in the symmetry-broken phase. This suggests the emergence of a second-order entanglement phase transition occurring at the same critical points as the dissipative phase transition. We perform a finite-size scaling analysis of the entanglement in Fig.~\ref{fig:bhd-ssb-ent}(c). Our results suggest that the maximum steady-state entropy achieved in the symmetry-broken phase tends to saturate with increasing $N$, while that in the symmetric phase vanishes as $N\to \infty$.

We also study the saturation time in the dynamics of the trajectory-averaged entanglement entropy $\overline{S_E}(\hat{a}_{+},\hat{a}_{-})$ and of the order parameter $\overline{\langle\hat{a}_{+}^\dagger\hat{a}_{+}\rangle}/N$. As shown in Fig.~\ref{fig:bhd-ssb-sat}(a) and (b), these two quantities follow very similar saturation dynamics. In the symmetric phase, both saturate very quickly at a time scale practically independent of $N$. In the symmetry-broken phase, they also feature fast saturation times. As shown in Fig.~\ref{fig:bhd-ssb-sat}(c), the saturation time of the entanglement entropy at $\tilde{F}=2.34\kappa$ (in the symmetric broken phase but closer to the left critical point~\footnote{The behavior at the right critical point is similar and not shown.}) scales as $\tau_{\mathrm{sat}}\sim \ln^p N$, with $p\simeq 3.2\pm0.1$ extracted from a fit. The order parameter $\overline{\langle\hat{a}_{+}^\dagger\hat{a}_{+}\rangle}/N$ shares essentially the same saturation time. In this regime, the experimental observation of the entanglement transition would not suffer from the exponential post-selection problem, but still requires a sub-exponential overhead.

\subsubsection{$J<0$: First-order phase transition with critical slowing down of entanglement}

Let us now consider the regime where the system admits a negative coupling constant. We still assume a vacuum initial state for the system, while adopting the following parameters: $J=-2.5\kappa$, $\Delta=-1.4\kappa$, and $U=2\kappa$.
The driven mode $\hat{a}_{-}$ is now the lower one in energy, and the drive frequency
is sufficiently blue-shifted with respect to this mode. Therefore, we expect the first-order dissipative phase transition corresponding to the optical bistability of the lower mode to dominate in the dynamics~\cite{garbinSpontaneousSymmetryBreaking2022}. 

We now perform the same analyses as for the previous regime. In Fig.~\ref{fig:bhd-bis-ent}(a) we recover the dissipative phase transition witnessed by the trajectory-averaged normalized total population $\overline{\langle\hat{n}_\mathrm{tot}\rangle}/N$. While the mean-field equations give two stable classical solutions (the higher and lower branches of the S-curve) for driving amplitudes close to the transition, the steady-state density matrix is unique since quantum fluctuations allow the state to tunnel between the two metastable solutions. This results in an average population between the high and the lower branches. In the thermodynamic limit $N\to\infty$, the steady-state population develops a first-order discontinuity between the low-density phase and the high-density one. 

Fig.~\ref{fig:bhd-bis-ent}(b) shows the trajectory-averaged steady-state entanglement entropy between the two spatial modes $\aaa_1$ and $\aaa_2$ across the dissipative phase transition discussed above. As $N$ increases, a first-order discontinuity also develops at the same critical point as the dissipative phase transition ($\tilde{F}\simeq 0.373\kappa$), followed by a peak in entanglement which could be due to quantum fluctuations driving the dissipative phase transition. Fig.~\ref{fig:bhd-bis-ent}(c) shows the scaling of the steady-state entanglement entropy as $N$ increases, in different phases. In the low-density phase, the entanglement saturates to a relatively low value. In the high-density phase close to criticality, the maximum entanglement entropy scales linearly with $\log N$. For larger driving values deep inside the high-density phase, the entanglement has a slower growth as a function of $N$ that is sub-logarithmic.

Finally, we study the transient dynamics of the order parameter $\overline{\langle\hat{n}_\mathrm{tot}\rangle}/N$ and of the trajectory-averaged entanglement entropy $\overline{S_E}(\aaa_1,\aaa_2)$. As shown in Fig.~\ref{fig:bhd-bis-sat}(a) and (b), these two quantities again follow very similar time dynamics. In both lower- and higher-density phases away from criticality, the relaxation time required to reach the steady state shows no significant dependence on $N$. At criticality, the dissipative phase transition associated with the optical bistability is known to display an exponential critical slowing down~\cite{casteelsCriticalDynamicalProperties2017}. This is indeed observed in the dynamics of the population $\overline{\langle\hat{n}_\mathrm{tot}\rangle}/N$ at a critical driving value $\tilde{F}=0.373\kappa$. The entanglement entropy is found to replicate this behavior. Fig.~\ref{fig:bhd-bis-sat}(c) shows the scaling of the saturation time $\tau_{\mathrm{sat}}$ of both the population and the entanglement at criticality, where both appear to follow the same exponential growth with $N$. We obtain a good fit to the exponential law $\tau_{\mathrm{sat}}\sim\rme^{pN}$ with $p=0.078\pm0.002$. Therefore, the entanglement phase transition of the Bose-Hubbard dimer in the optical bistability regime is clearly not post-selection-free in terms of its experimental detection, due to the critical slowing down in the entanglement dynamics.

\section{Experimental implementations}
\label{SecExp}

\begin{figure*}[t]
    \centering
    \includegraphics[width=\textwidth]{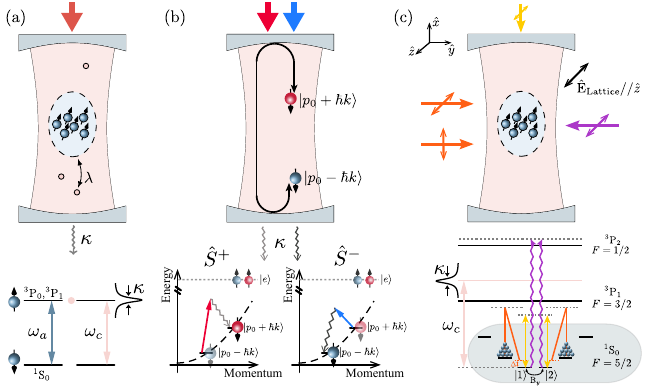}
    \caption{Proposed experimental implementations for (a) driven-dissipative Tavis-Cummings model, (b) driven spins with collective absorption and emission, and (c) driven-dissipative Bose-Hubbard dimer. (a) When the cavity mode $\omega_c$ is on resonance with the atomic transition $\omega_a$, the system realizes the Tavis-Cummings interaction. Working with narrow-linewidth transitions such as $^{1}\mathrm{S}_{0}\!-\,\!\!^{3}\mathrm{P}_0$ or $^{1}\mathrm{S}_{0}\!-\,\!\!^{3}\mathrm{P}_1$ on $^{88}$Sr and placing an additional resonant laser beam (red arrows) realizes the driven-dissipative Tavis-Cummings model. (b) We utilize two momentum states in the ground state of $^{87}$Rb as an effective two-level system (red and blue spin-down atoms). By shining two off-resonant laser beams (red and blue arrows) from the excited state $|e\rangle$, we can engineer and control the $\hat{S}^+, \hat{S}^-$ processes between the two-level system (bottom part). (c) We encode the ground state Zeeman levels $|m_F=-1/2\rangle$ ($|m_F=1/2\rangle$) of $^{173}$Yb as $| 1 \rangle$ ($| 2 \rangle$). The lattice polarization is parallel to the quantization axis along $\hat{z}$, providing a quadratic tensor shift to the ground state atoms. Off-resonant $\pi$-light from $^3\mathrm{P}_1$, sent through the cavity (yellow arrows), facilitates 4-photon processes that generate the local interactions. A transverse magnetic field $B_y$ provides the coherent coupling $J$ between $| 1 \rangle$ and $| 2 \rangle$. Two Raman beams from the side of the cavity (orange arrows) coherently populate $| 1 \rangle$ and $| 2 \rangle$ from $|m_F=\pm 3/2\rangle$ and also provide the detuning $\Delta$. An off-resonant beam (purple arrows) connecting to the $^3\mathrm{P}_2, F=1/2$ manifold engineers the dissipation of $| 1 \rangle$ and $| 2 \rangle$ via loss to the meta-stable states. }
    \label{figEric:Implementation}
\end{figure*}

In this section we describe implementations of all the models considered in the previous sections. We also use the case of the driven-dissipative Tavis-Cummings model to discuss in more detail the experimental aspects of postselection.

\subsection{Driven-dissipative Tavis-Cummings model}

As discussed in Sec.~\ref{sec:ddtc}, the driven-dissipative Tavis-Cummings model can be implemented naturally using cavity-QED. For definiteness, we describe here two possible implementations, both of them using Sr atoms.

In the first setup we consider an ensemble of $^{88}$Sr atoms within a high-finesse optical cavity and define a two-level system for each atom using the ground state $\ket{g}=\ket{^1\textrm{S}_0,m_J=0}$ and the long-lived excited state $\ket{e}=\ket{^3\text{P}_1,m_J=0}$. The collection of all these two-level systems constitutes the collective spin degree of freedom used in Eq.~(\ref{eqAnna:Ham_Dicke}). When the atoms are resonantly coupled to one of the cavity modes (frequency of $\ket{g}\leftrightarrow\ket{e}$ is equal to resonance frequency of the mode), the ensuing interaction between the mode and the atoms is naturally given by the Tavis-Cummings model $\propto \hat{a}^\dagger\hat{S}^-+\hat{a}\hat{S}^+$. The extra drive present in Eq.~(\ref{eqAnna:Ham_Dicke}) can then be realized by sending a laser, resonant with the $\ket{g}\longleftrightarrow\ket{e}$ transition, through the side of the cavity. Alternatively, the cavity can be driven directly instead, with no substantial modifications to the system properties. This experimental setting was indeed recently used to observe the phase diagram of the driven-dissipative Tavis-Cummings Model~\cite{song2024}. The steady state phases were determined by measuring the magnetization of the spins and  also the photons leaking out of the cavity.

If we use the parameters from Ref.~\cite{song2024,Muniz2021}, the cavity mode would have a full width at half maximum (FWHM) linewidth of $\kappa=2\pi\times153$~kHz (i.e. the lifetime of a photon inside the cavity is $1~\mu$s). The single photon Rabi frequency $2g=2\pi\times 21$~kHz reported therein is related to the $\lambda$ parameter in Eq.~(\ref{eqAnna:Ham_Dicke}) via $\lambda=g\sqrt{N}$. For moderate atom numbers $N=10^3-10^4$ this leads to $\lambda/\kappa=2.3-7.2$, very deep into the bistability regime. In this implementation, atoms in state $\ket{e}$ can also decay by spontaneously emitting photons into free space, with a $1/e$ decay time of $21~\mu$s. This process is not included in our model, so the results of Sec.~\ref{res_TC} are valid for times shorter than $21~\mu$s. To guarantee a clean observation of the driven-dissipative Tavis-Cummings dynamics, we can increase $\kappa$ (by e.g. reducing the cavity finesse) to accelerate the relevant timescales. A factor of $10$ increase in $\kappa$ would push the interaction strengths towards $\lambda/\kappa\approx 0.2-1.7$, which fall inside Fig.~\ref{figAnna_2} and still display bistability. Hence, current cavity-QED experiments can faithfully implement the driven-dissipative Tavis-Cummings model in the parameters regimes of interest.

The measurement process can be realized by using high efficency photodetectors to count the photons coming out through the cavity mirrors. Each detected photon would carry a time-tag, indicating when it was observed. Naively, repeating a quantum trajectory would then require repeating the same sequence of detections at exactly the same times. In practice, a coarse-graining procedure is necessary: the total evolution time is divided into intervals of size $\Delta t$ and detection events are binned into these intervals, whose length is chosen so that ideally at most one photon is detected within them. A given quantum trajectory is then defined by the presence (or absence) of a photon within each time interval. Repeating the same trajectory means repeating the same set of outcomes in each interval. From Appendix~\ref{app:binned_ev}, the time bins can be chosen to be as big as $\Delta t\approx \kappa^{-1}$ without substantially modifying the results of Sec.~\ref{res_TC}. Given that saturation times in the time-crystal phase [see Fig.~\ref{figAnna_5}(c)] are in the range $\kappa\tau_S=40-100$ this means, naively, that there are about $2^{40}-2^{100}$ possible trajectories, but some comments are in order:
\begin{itemize}
    \item Fig.~\ref{figAnna_5}(c) indicates that the dependence of $\kappa\tau_S$ with particle number is weak, with $\kappa\tau_S\approx A (\log N)^{\alpha}$. The large number of trajectories is instead a consequence of the prefactor $A$ being large. The favorable scaling with $N$ means that measurement strategies aimed at reducing the postselection overhead would be effective even for reasonably large numbers of particles.  
    \item The quantum state itself biases the trajectories that actually occur in a given trial, meaning that not all trajectories are explored. This point is made clearer by considering the cavity to have an average number of photons $\ll 1$ (e.g. if the system is driven along the cavity direction~\cite{song2024}) or if the time bins are chosen to be very small. Then a given time bin will report no detection across many repetitions of the conditional dynamics.
    \item The saturation time $\tau_S$ may also be a memory time~\cite{lin2023}, indicating that outcomes in time bins far in the past matter less, reducing the effective number of bins that need to be tracked.
\end{itemize} 

\begin{figure}
    \centering
    {\begin{overpic}
    [width=0.485\linewidth]{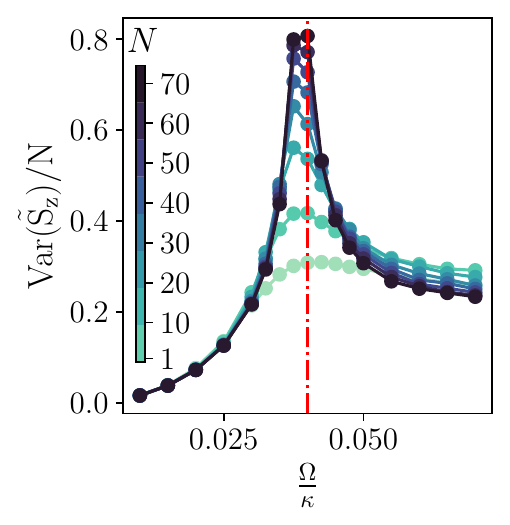} \put(0, 90) {(a)}
    \end{overpic}}
    {\begin{overpic}
    [width=0.485\linewidth]{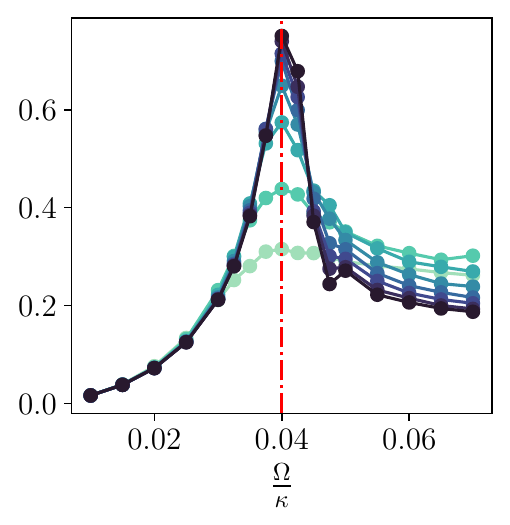} \put(3, 90) {(b)}
    \end{overpic}}
    \caption{Phase diagrams obtained from the size-rescaled variance $\hat S_z$. All the plots are obtained with $N_{\rm{traj}}=4000$ trajectories at $\lambda/\kappa=0.2$. The plots correctly signal the monitored critical point $\Omega_C/\kappa=0.04$ and share the legend. (a) Phase diagram obtained in the steady state. (b) Phase diagram obtained respectively at $\kappa t=110$ smoothing the oscillations in the entangled phase with a Gaussian filter with variance equal to the oscillations' period.}
    \label{figAnna:vars}
\end{figure}

The monitored transition can also be probed using other non-linear quantities in the system state that are easier to measure than the entanglement entropy~\cite{Moghaddam}. The variance of $\hat S_z$, which is easier to access in a given quantum trajectory, clearly signals the monitored transition with a peak at the correct critical point, as shown in Fig. \ref{figAnna:vars}(a). The phase diagram in Fig. \ref{figAnna:vars} is obtained at $\lambda/\kappa=0.2$ from the steady state values of the dynamics (extracted at $\kappa t \sim 500$), similarly to what was done previously for the entanglement entropy. Note that the transition can be witnessed likewise at earlier times. [Panel (b)] shows the variance of $\hat S_z$ obtained at $\kappa t=110$ when saturation has not yet been reached; the critical point is still well-signaled by a peak. The saturation time generally depends significantly on the quantity used to probe the transition. However, the possibility of observing the transition at earlier times could help reduce the required number of trajectories.

All of these considerations reduce the postselection overhead, but we leave an in-depth investigation of them for future work.

The second implementation uses instead  $^{87}$Sr. We can define the two-level system using the ground states $\ket{g}=\ket{^1\textrm{S}_0}$ and the ultra long-lived excited states $\ket{e}=\ket{^3\text{P}_0}$. Using the same cavity with a FWHM linewidth $\kappa=2\pi\times153$~kHz, a single photon Rabi frequency of $2g=2\pi\times8$~Hz~\cite{Matthew}, and $N=10^4-10^5$ leads to $\lambda/\kappa=0.001-0.01$. This is very close to the horizontal axis in Fig.~\ref{figAnna_2}, but still displays a transition between a stationary phase and a time crystal phase (see the discussion after Eq.~(\ref{eqAnna:MasterEq_Dicke})). In this limit the ensuing dissipative/measurement dynamics occurs with the timescale $\lambda^2/\kappa$ ($15-150$~ms)~\cite{Carmichael,Iemini} and the average time between photon detections increases. The size of the time bins can thus be made to be $\Delta t\sim \lambda^2/\kappa$, much larger than $\kappa^{-1}$. Furthermore, spontaneous emission into free space now has a $1/e$ decay time $\sim 100$~s, and is thus not relevant when compared to $\lambda^2/\kappa$.

\subsection{Driven spins with collective absorption and
emission}
Besides the approach described in Sec.~\ref{sec:dscae}, which is based on coupling an ensemble of two-level emitters to a finite temperature cavity, we provide here another implementation based on the experiment done in Ref.~\cite{luo2024}. In this setup there is an ensemble of $^{87}$Rb atoms inside an optical cavity, with no spatial confinement along the cavity axis. The two-level system is then defined in terms of motional states, with $\ket{\downarrow}=\ket{p_0-\hbar k}$ and $\ket{\uparrow}=\ket{p_0+\hbar k}$ representing wavepackets with well-defined momenta. The collective dissipation described by Eq.~(\ref{eq:btc-finite-t-lindblad}) can be generated via four-photon Raman processes involving the cavity and two assisting lasers. The resulting jump operator 
\begin{equation}\label{eqn:ExperimentalImplemtaton:JumpOperator}
    \hat{L}\propto \beta_1\hat{S}^-+\beta_2 \hat{S}^+ e^{i\delta t}
\end{equation}
is determined by parameters $\beta_{1,2}$ and $\delta$ that can be controlled by changing the power and frequency of the assisting lasers. When $\delta$ is larger than other timescales of the system, this leads to independent collective decay ($\hat{S}^-$) and absorption ($\hat{S}^+$) processes. The single particle coherent drive can then be implemented using two-photon Bragg pulses~\cite{Greve2022}.

The quantum jumps associated to Eq.~(\ref{eqn:ExperimentalImplemtaton:JumpOperator}) are monitored via photodetection of the light escaping through the cavity mirrors. Each detected photon mathematically corresponds to the application of the operator $\hat{L}$ on the atomic ensemble. The distinction between $\hat{S}^+$ and $\hat{S}^-$ is then encoded in the frequency of the detected photon, provided the average time between each photon is large enough that frequencies of size $\delta$ can be resolved (easier to do at large $\delta$). Although such a frequency-selective photodetection may be challenging to implement, heterodyne detection of the output light would distinguish the $\hat{S}^-$ and $\hat{S}^+$ processes and might not modify substantially the results of section~\ref{res_finiteTSpin} (see Ref.~\cite{Passarelli} for a related discussion). Alternatively, proposals that use more than one cavity mode could be explored~\cite{shimshi2024}. In these alternative setups, the $\hat{S}^+$ and $\hat{S}^-$ processes are coupled to different cavity modes that can be unambiguously differentiated by e.g. frequency, or spatial structure, or the modes could be associated to different cavities. This allows for easier discrimination between the different quantum jump processes.

\subsection{Driven-dissipative Bose-Hubbard dimer}\label{DBH}
Here we provide two implementations of the driven-dissipative Bose-Hubbard dimer. The first implementation is in cavity-QED using  also alkaline-earth atoms but here we  consider instead an ensemble of $^{173}$Yb atoms trapped in a one dimensional lattice inside an optical cavity. The ground state of $^{173}$Yb ($^1S_0$) consists of $6$ levels ($F=5/2$), and we define the quantization axis ($\hat{z}$) to be parallel to the lattice polarization, which is transverse to the cavity axis ($\hat{x}$, see Fig~\ref{figEric:Implementation}(c)). We choose $\ket{1}\equiv\ket{m_F=-1/2}$ and $\ket{2}\equiv \ket{m_F=1/2}$ to encode the two bosonic modes analyzed in Section~\ref{sec:bhd}, such that $\hat{a}_{1}^\dagger\hat{a}_1$ and $\hat{a}_2^\dagger\hat{a}_2$ (as defined in Section~\ref{sec:bhd}) correspond to the number of atoms in $\ket{1}$ and $\ket{2}$, respectively. Importantly, the $\hat{a}_{1,2}$ are bosonic because they encode symmetric permutations among the spin states of the Yb atoms, although the isotope in question is fermionic. The tensor light shift caused by lattice light leads to a parabolic profile for the ground state energies (as shown in Fig.~\ref{figEric:Implementation}(c)), so $\ket{1}$ and $\ket{2}$ are isolated from the other ground states. 

The tunneling $-J(\hat{a}_1^\dagger\hat{a}_2+\hat{a}_2^\dagger\hat{a}_1)$ can be engineered  by adding a transverse magnetic field (perpendicular to the lattice polarization), as shown in Fig.~\ref{figEric:Implementation}(c). Alternatively, it can be engineered using a single laser tone, propagating in the $y$ direction and circularly polarized, to generate two-photon Raman transitions mediated by states in the excited $^3P_1$ ($F=3/2$) manifold. The coherent drives $F_i$ are implemented via another set of Raman pulses (orange arrows) that coherently couple $\ket{1}$ with $\ket{m_F=-3/2}$ and $\ket{2}$ with $\ket{m_F=3/2}$, respectively. When  preparing a large number of atoms in $\ket{m_F=\pm3/2}$  they act as particle reservoirs. The two-photon detuning will then control the $\Delta$ in Eq.~(\ref{eq:bh-ham}), or alternatively the strength of the axial field that set the quantization axis. 

Interactions are generated by dispersively coupling the $^1S_0\leftrightarrow\!^3P_1$ optical transition to a far-detuned optical cavity mode and driving the cavity with light (yellow arrows) that is moderately detuned from the $^1S_0\leftrightarrow\!^3P_1$ transitions. This generates correlated four-photon processes where a ground state atom scatters a drive photon into the cavity, while another ground state atom scatters that cavity photon back to the drive~\cite{Chu2023}~(see Appendix~\ref{app:ExpImp}). Besides the onsite interactions, this scheme also generates density-density terms of the form $\hat{a}_1^\dagger\hat{a}_1\hat{a}_2^\dagger\hat{a}_2$. The latter could be canceled for appropriate parameters by using the contact interaction between the $|m_F=\pm1/2\rangle$ nuclear spin levels. While the scattering length can not be easily tuned via Feshbach resonance, the use of Ramam drives to couple the ground states to the $^3P_0$ states can open some degree of tunability. We leave for future work a more in-depth investigation of the effect of this new interaction on the results of Section~\ref{sec:result-bose-hubbard}.

Finally, to mimic the photodetection process we send laser light polarized along $z$, off-resonant with the $^1S_0\leftrightarrow\!^3P_2~(F=1/2)$ transition (purple arrows), so that atoms in $\ket{1}$ and $\ket{2}$ get incoherently excited to the $^3P_2$ states. Since $^3P_2$ are meta-stable excited states, atoms that end up in these excited states are ``lost" from the dynamics in the ground state manifold, giving rise to $\hat{a}_{1,2}$ jump operators. To monitor the number of atoms that have been lost, we couple the $^3P_2~(F=1/2)\leftrightarrow\!^3D_3~(F=3/2)$ transition to another cavity mode and perform continuous non-demolition readout of atom number by keeping track of the cavity frequency shift.

The second (modified) implementation is with circuit-QED platforms, where dissipative Bose-Hubbard models have been demonstrated in the past~\cite{Ma,Carusotto2020}. Following Ref.~\cite{Ma}), two transmon qubits, acting as nonlinear microwave resonators, constitute sites $1$ and $2$ in Section~\ref{sec:bhd}. The tunnelling $J$ can be created by capacitively coupling the resonator and is fixed by fabrication, but the onsite energies $\Delta$ can be independently controlled. The interaction energy $U$ is similarly fixed by the fabrication procedure and generally negative. The parameters reported in Refs.~\cite{Ma,Saxberg2022,Roberts2024} correspond to large $U/J\approx 15-25$ but it should be possible to reduce this ratio. Alternatively, a tunable $J$ can be engineered by coupling the transmons via a three-wave mixing element such as a SNAIL~\cite{Frattini2017}. This allows for a tunable $J$, controlled by keeping the resonators spectrally separated and externally driving the SNAIL at the respective frequency difference. Furthermore, in this configuration the two transmons could be replaced by resonators, and would inherit weak nonlinearities from their coupling to the three-wave element, allowing for ratios $J/U$ closer to the values inspected in Section~\ref{sec:result-bose-hubbard}. 

The drives $F_i$ are generated by coupling each resonator to a transmission line and sending microwave with variable strengths. The measurement process is then implemented by monitoring the light leaking out from the resonators into their respective lines. Although high-efficiency single microwave photon detectors are not commercially available, single microwave photon detection has been demonstrated in the past~\cite{Opremcak2018,Albertinale2021}. Furthermore, it is technically more convenient to perform homodyne detection of the outgoing light, and it would thus be interesting to consider how the results of Section~\ref{sec:result-bose-hubbard} are modified. We leave this to future work.

\section{Conclusions}\label{SecIV}

In this paper, we performed a thorough investigation of monitored dynamics and phase transitions in infinite-range systems. We introduced and analyzed three collective monitored models: the Tavis-Cummings model, the Superradiance model, and the Bose-Hubbard dimer. Our goal was to study their monitored dynamics and steady-state properties, focusing particularly on measurement-induced phase transitions in entanglement entropy. For all models, we observed that the entanglement transitions occur at the same critical point as the already-known phase transitions in the mixed state. The non-trivial dynamics and mixed-state phase diagrams provided deeper insights into these phase transitions. Specifically, we studied whether these entanglement phase transitions exhibit fast-saturating dynamics, which would reduce the post-selection overhead due to their semiclassical nature.

In the case of the driven-dissipative Tavis-Cummings model, our results suggest that the entanglement phase transition is subject to a post-selection barrier that scales polynomially with system size for values of the light-matter interaction $\lambda/\kappa$ outside the bistable region. When the bistable region is met, however, we find that the thermodynamic limit and the infinite-time limit no longer commute, leading to a slow saturation of the entanglement entropy. This results in a monitored phase transition that suffers from a significant post-selection challenge.

The Superradiance model exhibits unique dynamics determined by the relative weights of the two jump operators, $\hat{S}^+$ and $\hat{S}^-$, which can be interpreted as temperature-like effects in the dissipator. Although the mean-field equations remain temperature-independent, the entanglement transition's scaling in size in the steady state is modified by temperature. However, the saturation time does not depend on temperature, and the monitored transition preserves the polynomial post-selection problem already found at $T=0$ \cite{Passarelli}.

Similarly to the Tavis-Cummings model, we examined the Bose-Hubbard dimer under two conditions. For positive values of the cavity-cavity coupling, $J$, the entanglement phase transition is characterized by fast-saturating entanglement, resulting in a polynomial post-selection problem. In contrast, for negative values of $J$, where the system is known to exhibit bistability and undergoes a first-order phase transition in the mixed state, the monitored dynamics also exhibits slow saturation times, leading to high post-selection overhead.

Our study suggests that semiclassical dynamics alone does not guarantee fast-saturating entanglement or reduced post-selection overhead. When bistability and first-order phase transitions are present, the associated critical slowdown also affects entanglement entropy, hindering the favorable post-selection properties observed in other collective models.

Considering the regions with favorable post-selection properties, we propose experimental implementations of our monitored models. This discussion allowed us to identify additional challenges in observing these monitored phase transitions, including the impact of finite time resolution of the devices and the need to reduce the numerical coefficients that define the saturation time. In this context, we propose several considerations, including the role of typicality and memory effects in quantum trajectories, and the possibility of observing transitions at earlier times to minimize post-selection overhead.

Lastly, we note that the conclusions presented here are specific to monitored phase transitions in entanglement entropy. Such transitions could also be probed using other non-linear observables~\cite{Moghaddam}, such as the spin variances  as shown in this paper, but the properties of the phase transition may vary depending on the choice of observable. This remains an open question for future studies.

\section*{Acknowledgments}
We would like to thank C.\,Ciuti, L.\,Correale, V.\,Heyraud, P.\,Lucignano, A.\,Russomanno, M.\,Schir\`o, C.\,Vanoni and J.\,Venkatraman for very useful discussions. G.\,P. acknowledges computational resources
from the CINECA award under the ISCRA initiative, and from MUR, PON “Ricerca e Innovazione 2014-2020”, under Grant No.\,PIR01\_00011
- (I.Bi.S.Co.). D.B. acknowledges support from a Simons Investigator Award (Grant No. 511029). This work was supported by PNRR MUR project PE0000023- NQSTI, by the European Union (ERC, RAVE, 101053159). Views and opinions expressed are however those of the author(s) only and do not necessarily reflect those of the European Union or the European Research Council. Neither the European  Union nor the granting authority can be held responsible.

\bibliography{BiblioEPT_Semi}

\appendix

\section{Dissipative-Driven Tavis-Cummings Model}\label{app:TavisCummings}
The behavior of the Driven-Dissipative Tavis-Cummings Model has been studied in the thermodynamic limit through mean-field equations for the spins' magnetizations and the field's quadratures~\cite{Mattes}. To this stand, it is important to introduce the normalized operators $\hat m_j = 2\hat S^j/N$ with $j=x,y,z,\pm$, $\hat m_q = (\hat a^\dag + \hat a)/\sqrt{2N}$ and $\hat m_p = i(\hat a^\dag - \hat a )/\sqrt{2N}$.
These operators are well-defined in the thermodynamic limit and admit vanishing commutators for $N\to\infty$, which justifies the mean-field decoupling $\expval{\hat m_i \hat m_j}=\expval{\hat m_i}\expval{\hat m_j} \equiv m_i m_j$~\cite{Carollo} in the semiclassical picture.

The average values of the magnetizations and quadratures evolve according to Heisenberg equations given by the adjoint master equation~\cite{breuer2002theory}.  Setting $\expval{\hat m_p(0)}=0$ and $\expval{\hat m_x(0)}=0$, one simply can study:
\begin{eqnarray}\label{eqAnna:MF}
    \dfrac{\d{m}_y}{\d t} & = & -2\Omega m_z -\sqrt{2}\lambda m_q m_z \nonumber, \\
    \dfrac{\d{m}_z}{\d t} & = & 2 \Omega  m_y +\sqrt{2}\lambda  m_q  m_y, \\ 
    \dfrac{\d{m}_q}{\d t} & = & -\frac{\lambda}{\sqrt{2}}  m_y - \frac{\kappa}{2} m_q \nonumber,
\end{eqnarray}
where $m_\mu=\expval{\hat m_\mu}$.\\
The semiclassical dynamics of the model can be studied through (deterministic) trajectories characterized by the evolution of the expectation values $(m_y(t),m_z(t),m_q(t))$. Due to conservation laws, the dynamics we are studying occurs on the surface of the cylinder given by $m_y^2+m_z^2=1$ and $m_q\in\mathbb R$.

Based on the type of trajectories, three phases can be identified as in Fig.~\eqref{figAnna_2}(a) as already explained in the main text: a stationary phase, a time crystal phase, and a bistable phase. The boundary of the time-crystal phase can be derived analytically: $\Omega_C/\kappa=\lambda^2/\kappa$. The boundary between the stationary phase and bistability phase can be only determined numerically by checking the uniqueness of the steady-state from several random initial conditions. It corresponds to the line with error bars in Fig.~\ref{figAnna_2}. Notice that the phase diagram presents a tricritical point (red dot), above which the bistability region appears.

The difference between stationary behavior and time crystal behavior can be characterized by the magnetization order parameter:
\begin{equation}\label{eqAnna:opDicke}
    \tilde m_z = \lim_{t_f\to\infty}\frac{1}{t_f}\int_{0}^{t_f}dt \, m_z(t),
\end{equation}
that corresponds to the average value over time of the magnetization $m_z(t)$.
Solving the differential equations~\eqref{eqAnna:MF} one can easily realize that the stationary behavior yields $\tilde m_z\le0$, while the time-crystal limit cycle behavior attains positive values for the order parameter~\cite{Mattes, Cabot}. Notice that the order parameter is blind to the bistability region. In this phase, $\tilde m_z$ can only detect the type of behavior given by the chosen initial state.

Through this order parameter, a continuous phase transition between stationary and time-crystal phases can be witnessed below the tricritical point. Above the tricritical point, $\tilde m_z$ detects a first-order phase transition between stationary and time crystal behavior that happens in the bistable phase with a critical point determined by the initial condition.

\section{Finite-temperature spin model}\label{app:FiniteTSpin}

As for all collective models, we can easily derive the mean-field (MF) equations of motion of the magnetization components $\vec{m}$ starting from the Ehrenfest equations and exploiting the fact that correlations become negligible in the thermodynamic limit so that the equations can be closed at the first order in the cumulant expansion ($\langle \hat m_i \hat m_j \rangle = m_i m_j$). The surprising result, already noted~\cite{carollo2023}, is that the MF equations are independent of the bath temperature $T$: at the mean-field level, which is exact for large $N$, the finite-$T$ superradiant model behaves exactly as its zero-temperature limit. For collective systems, the proper high-temperature limit entails rescaling the temperature with the system size $N$ so to have a sizeable effect when $N\to\infty$~\cite{souza2022}.

Let us briefly compare the two scenarios. Without any temperature rescaling, the MF equations of motion read
\begin{subequations}\label{eq:btc-finite-t-mf}
    \begin{align}
        \dfrac{\d m_x}{\d t} &= -\kappa m_x m_z \, ,\\
        \dfrac{\d m_y}{\d t} &= -\Omega m_z - \kappa m_y m_z \, ,\\
        \dfrac{\d m_z}{\d t} &= \Omega m_y + \kappa (m_x^2 + m_y^2),
    \end{align}
\end{subequations}
which, as anticipated, do not depend on $\beta$. If, on the other hand, we scale the temperature with $N$ according to the scaling law $\beta \to \beta / N$, the equations of motion become
\begin{subequations}\label{eq:btc-finite-t-mf-rescaled}
    \begin{align}
        \dfrac{ \d m_x}{\d t} &= -\kappa m_x m_z - \frac{2 \kappa}{\beta} m_x \, ,\\
        \dfrac{ \d m_y}{\d t} &= -\Omega m_z - \kappa m_y m_z - \frac{2 \kappa}{\beta} m_y \, ,\\
        \dfrac{ \d m_z}{\d t} &= \Omega m_y + \kappa (m_x^2 + m_y^2) - \frac{4\kappa}{\beta} m_z.
    \end{align}
\end{subequations}
Compared to Eqs.~\eqref{eq:btc-finite-t-mf}, in each of Eqs.~\eqref{eq:btc-finite-t-mf-rescaled} there is an extra term, proportional to the component itself and to the temperature $T$. Terms like these also arise when local dissipation is added to the Lindbladian of Eq.~\eqref{eq:btc-finite-t-lindblad}~\cite{paz2021}. The effect of local dissipation is to destroy the time-crystal dissipative phase transition~\cite{passarelli2022} and make the system always converge to the featureless fixed point $m_i = 0$.

Therefore, in the main text of this manuscript, we consider the case where the temperature is not rescaled with $N$.

\section{Driven-Dissipative Bose-Hubbard Dimer}\label{app:BHDimer}
We briefly discuss in this appendix the mean-field equations of the Bose-Hubbard dimer.
In a mean-field approximation, where one assumes the factorization $\langle\aaa_i^{\dagger m}\aaa_j^{n}\rangle\simeq \langle\daaa_i\rangle^m\langle\aaa_j\rangle^n$, the fields $\alpha_i\equiv\langle\aaa_i\rangle$ evolve under the following dynamics:
\bea\label{eq:mf-bhd}
    \dfrac{\d \alpha_i}{\d t} = \rmi\left(\Delta\alpha_i-U\abss{\alpha_i}\alpha_i-F_i+J\alpha_{i'}\right)-\dfrac{\kappa}{2}\alpha_i\,,
\eea
where $i'$ denotes the site that is not $i$ within the dimer. 

 Note that the mean-field equation~\eqref{eq:mf-bhd} is invariant if we scale the driving $F$, the nonlinearity $U$ and the field amplitude $\alpha_i$ simultaneously according to
 \bea
 U\longrightarrow\tilde{U}/N\,,\quad F_i\longrightarrow\sqrt{N}\tilde{F}_i\,,\quad \alpha_i\to \sqrt{N}\alphatil_i\,.
 \eea
 The limit $N\to\infty$ is therefore a well-defined thermodynamic limit for the Bose-Hubbard model.

\section{Saturation time}\label{app:sat}
The saturation time $\tau$ corresponds to the time needed to reach the steady state value for the operator $\hat o(t)$. We are interested in particular in the saturation time of the magnetization and the entanglement entropy. In general, since the average value of the magnetization is linear in the density matrix ($\expval{\hat m(t)}=\Tr{\hat m\hat\rho(t)}$), the saturation time of the magnetization $\tau_m$ corresponds to the real part of the longest-lived eigenvalue of the Liouvillian. This procedure can be performed only for very small system sizes, thus we obtained $\tau_m$ numerically. Also the saturation time of the entanglement entropy $\tau_S$ has been obtained numerically due to its less trivial dependence on the density matrix. 

We obtained the saturation time for the magnetization $\tau_m$ and the entanglement entropy $\tau_S$ from the data shown in Figs.~\ref{figAnna_4},~\ref{figAnna_6} and~\ref{figAnna_8}. 
The procedure requires smoothing out the oscillations of the time traces of $\expval{\hat o(t)}$ with a Gaussian filter and fitting the smoothed curves with an exponential decay $F_{\mathrm{fit}}(t) = a e^{-t/b} +c$ \cite{Passarelli}. In many cases, to obtain a better fit, it has been necessary to fit the function $|\expval{\hat o(t)}-o_{\mathrm{SS}}|$ with $o_{\mathrm{SS}}$ being the steady state value of the observable. The choice of fitting with a decaying exponential is due to the knowledge of the evolution of the density matrix:
\begin{equation}
    \ket{\rho(t)}\!\rangle = e^{\mathcal{L}t}\ket{\rho(0)}\!\rangle = \sum_{\mu=0}^{\mathrm{dim} \mathcal H^2 -1}\langle\!\bra{\pi_\mu}\ket{\rho(0)}\!\rangle \, e^{\lambda_\mu t} \ket{\rho_\mu}\!\rangle\, .
\end{equation}
An estimate of the saturation time can be obtained from the $b$ parameter of the fit.

\begin{figure}
    \centering
    \includegraphics[scale=0.9]{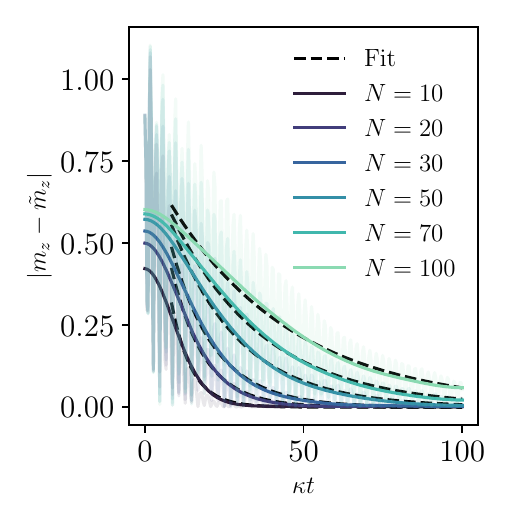}
    \caption{Smoothing of the magnetization curves for $\lambda/\kappa=0.8$ and $\Omega/\kappa=0.8$ for different sizes, and their exponential fit (dashed lines). The opaque solid lines correspond to the original oscillating function.} 
    \label{figAnna_14}
\end{figure}

An example of the procedure is shown in Fig.~\ref{figAnna_14} for the magnetization $\tilde m_z$ in the upper part of the phase diagram. The plot is obtained for $\Omega/\kappa=\lambda/\kappa=0.8$ and shows the original curves (opaque solid lines), the smoothed-out curves (solid lines), and the exponential fit (dashed lines).

\section{Time-Resolution effects on the entanglement transition}\label{app:binned_ev}
Throughout this paper, we have shown in which cases measurement-induced phase transitions in infinite-range systems admit a fast-saturating entanglement and can be reproduced experimentally with a polynomial post-selection problem. 
The experimental realization of the physics we are presenting, however, has an intrinsic challenge compared to numerical simulations: the fact that experimental devices have a finite resolution $\Delta \mathscr T$, which does not allow to distinguish trajectories that differ within time intervals $\delta t< \Delta \mathscr T$. For example, two trajectories having a jump at different jump times inside the interval $[t,t+\Delta \mathscr T]$, may not be distinguishable in experimental implementations. Not knowing the precise occurrence of quantum jumps implies partial information about the quantum state, hence the quantum state is no longer pure.
In practice, this means that the building block of our simulations, the pure-state-trajectory, is replaced in experiments by a mixed-state-trajectory corresponding to the average of pure-state-trajectories that differ only inside bins. Suppose $\left\{\ket{\psi_{\gamma_1}},...,\ket{\psi_{\gamma_n}}  \right\}$ is a family of pure-state-trajectories which are indistinguishable from an experimental point of view. The associated mixed-state-trajectory is defined as
\begin{equation}
    \hat \rho_\gamma = \frac{1}{n}\sum_{i=1}^n \ket{\psi_{\gamma_i}}\bra{\psi_{\gamma_i}}
\end{equation}
 
This inaccuracy in the trajectories' resolution does not affect average observables but does modifies those quantities that are non-linear in the state. For this reason, the question we would like to answer concerns how the experimental resolution $\Delta \mathscr T$ affects the transition we are witnessing from the exact simulations. In the case of entanglement entropy, the question translates to understanding how the binned time evolution decreases the purity of $\hat\rho_\gamma$ and whether the change in the behavior of the (entanglement) entropy reflects the behavior of quantum correlations or classical correlations.

We tackle this problem by writing a conditional master equation for an approximate time evolution of $\hat  \rho_\gamma$. This allows us to estimate the (entanglement) entropy in the steady state and verify how the scaling with the system size changes. The procedure corresponds to a rewriting of the quantum jump master equation considering the evolution at fixed time steps $t_i\in(0,\Delta \mathscr T, 2\Delta \mathscr T, ...., t_f)$ of the mixed state $\hat\rho_\gamma$.

Consider now a system described by a Hamiltonian $\hat H_0$. We monitor events associated with the quantum jump $\hat L$ occurring with rate $\kappa$. Our discussion is trivially generalized to the case of multiple dissipative channels. Restricting to the case in which the waiting time is comparable with the experimental resolution $\Delta \mathscr T$, we can assume to have only zero jumps or one jump inside a bin.

We label as $\hat \rho_\gamma(t_i)$ the state of the system at time $t_i$, and as $\hat{\mathcal U}_{\Delta\mathscr T}=\text{exp}\{-i\Delta \mathscr{T} (\hat H_0-\frac{i\kappa}{2} \hat L^\dag \hat L)\}$ the non-unitary evolution operator corresponding to non-detection. \\The evolved state at time $t_{i+1}$ is given with probability $p_0=1-\kappa\Delta \mathscr T\expval{\hat L^\dagger \hat L}_{t_i}$ by
\begin{equation}\label{eq_binme_nj}
    \hat \rho_\gamma(t_{i+1})^{0\rm{J}} = \frac{\hat{\mathcal U}_{\Delta\mathscr T}\, \,  \hat \rho_\gamma(t_i) \, \, \hat{\mathcal U}_{\Delta\mathscr T}^\dag}{\Tr{\hat{\mathcal U}_{\Delta\mathscr T}^\dag \hat{\mathcal U}_{\Delta\mathscr T} \,  \hat \rho_\gamma(t_i)}},
\end{equation}
in the non-detetction case, or with probability $p_1=\kappa\Delta \mathscr T\expval{\hat L^\dagger \hat L}_{t_i}$ by
\begin{eqnarray}\label{eq_binme_j}
    &&\hat \rho_\gamma(t_{i+1})^{1\rm{J}} = \\=&&\int_0^1 \! \frac{\hat{\mathcal U}_{x\Delta\mathscr T}\, \hat a \,  \hat{\mathcal U}_{(1-x)\Delta\mathscr T} \, \hat \rho_\gamma(t_i) \,\hat{\mathcal U}_{(1-x)\Delta\mathscr T}^\dag \, \hat a^\dag \, \hat{\mathcal U}_{x\Delta\mathscr T}^\dag}{\Tr{\hat{\mathcal U}_{x\Delta\mathscr T} \, \hat a \, \hat{\mathcal U}_{(1-x)\Delta\mathscr T} \, \hat \rho_\gamma(t_i) \,\hat{\mathcal U}_{(1-x)\Delta\mathscr T}^\dag \, \hat a^\dag \, \hat{\mathcal U}_{x\Delta\mathscr T}^\dag}} \, dx \nonumber .
\end{eqnarray}
in the detection case.
The last expression takes into account the fact that the jump may happen at any time inside the bin and calculates the average over all the possibilities.

A further approximation has been introduced: we are defining the probability of the jump event based on the density $\expval{\hat L^\dag \hat L}$ calculated at the beginning of the bin, while it may change due to the intra-bin evolution. We assume the approximation to hold since the resolution should be smaller than the timescale of the dynamics in all reasonably accurate experimental setups.

\subsection{The Tavis-Cummings Model}

We apply this method to the Tavis-Cummings model, focusing on the region with $\lambda/\kappa=0.2$ where the mitigation of the post-selection problem applies.

\begin{figure}
    \centering
    {\begin{overpic}
    [width=0.48\linewidth]{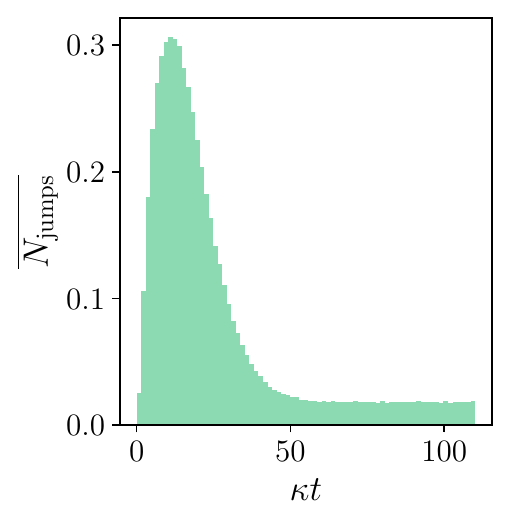}
    \put (0, 85) {(a)}
    \end{overpic}}
    {\begin{overpic}
    [width=0.48\linewidth]{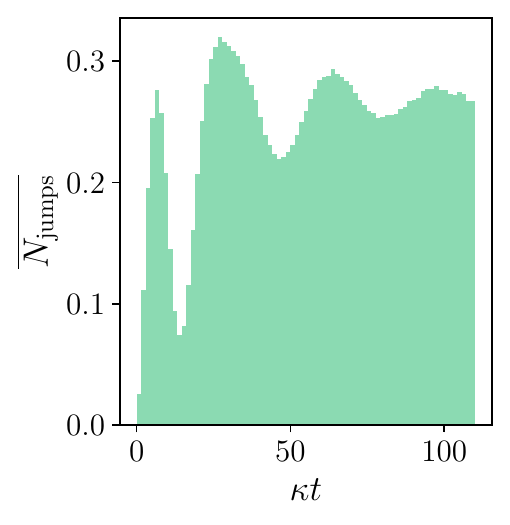}
    \put (-5, 85) {(b)}
    \end{overpic}}
    \caption{(a) and (b): average number of jumps for the binned evolution respectively in the stationary and time crystal phase. The bin width is $\Delta \mathscr T=1.4/\kappa$ and the statistics has been obtained from 120 000 pure-state-trajectories for $N=5$.} 
    \label{figAnna_app1}
\end{figure}

\begin{figure}
    \centering
    {\begin{overpic}
    [width=0.48\linewidth]{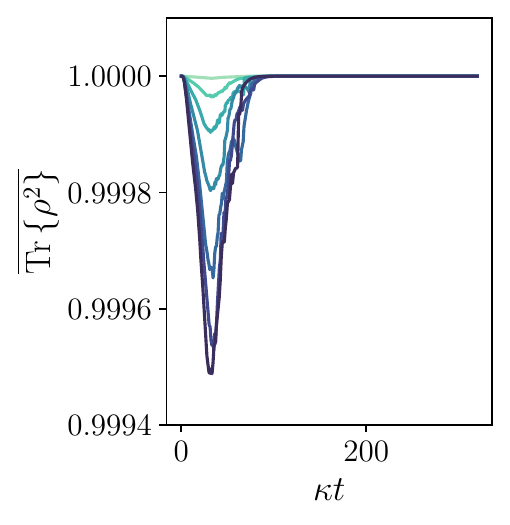}
    \put (0, 85) {(a)}
    \end{overpic}}
    {\begin{overpic}
    [width=0.48\linewidth]{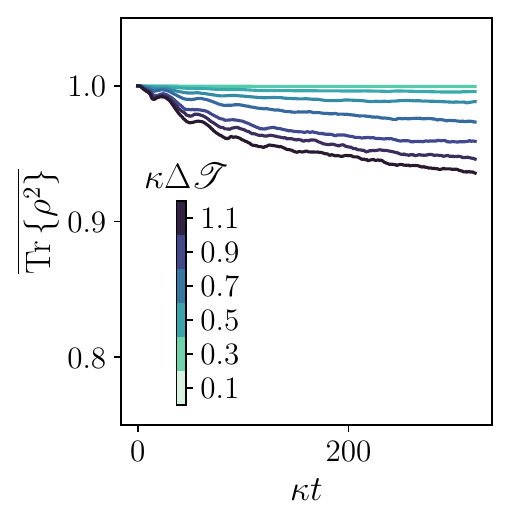}
    \put (-5, 85) {(b)}
    \end{overpic}}
    \caption{(a) and (b): Purity evolution respectively in the stationary and time crystal phase for several binning choices. The average is obtained from 1000 mixed-state trajectories for $N=9$. The legend is shared.} 
    \label{figAnna_app2}
\end{figure}

\begin{figure}
    \centering
    {\begin{overpic}
    [width=0.48\linewidth]{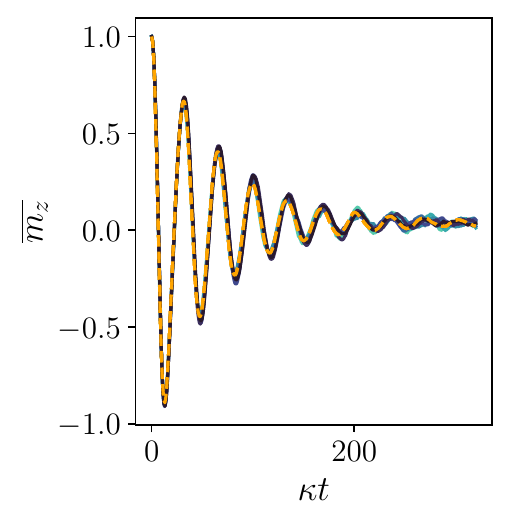}
    \put (0, 85) {(a)}
    \end{overpic}}
    {\begin{overpic}
    [width=0.48\linewidth]{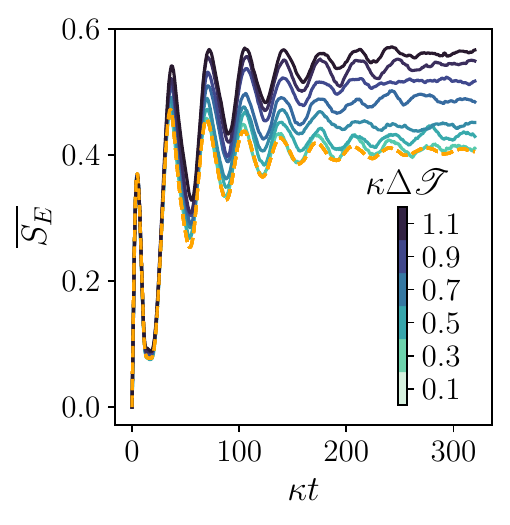}
    \put (-5, 85) {(b)}
    \end{overpic}}
    \caption{(a) Magnetization dynamics for several bin sizes compared to the exact result (orange dashed line). (b)  Entanglement dynamics for several bin sizes compared to the exact result (orange dashed line). For both plots the average is obtained from 1000 mixed-state trajectories for $N=9$. The legend is shared and both plot represent the dynamics in the time-crystal phase.} 
    \label{figAnna_app24}
\end{figure}

We repeat the simulation for several choices of bin widths, each corresponding to a different time resolution of a putative experimental setup. The bin widths we choose cover the range $\Delta \mathscr T\in [0.1/\kappa, 1.1/\kappa]$. The statistics of jumps has been studied for each binning (an example for $N=5$ is shown in Fig.~\ref{figAnna_app1}) and for each bin and size the condition $\kappa\Delta \mathscr T \expval{\hat a^\dag \hat a}<1$ is verified, meaning that on average the number of jumps is less than one inside each bin and the proposed master equation represents a good approximation.
Furthermore, to confirm the quality of the approximation, we show that the purity does not decrease dramatically in Fig.~\ref{figAnna_app2} justifying the calculation of the entanglement entropy as if the state were pure. Interestingly, notice how the purity of the stationary phase in [panel (a)] can be safely considered to be always 1, indicating this phase to be stable with respect to the choice of binning. 

As anticipated, observable averages do not depend on the binning procedure as the trajectory and quantum average commute. [Panel (a)] of Fig.~\ref{figAnna_app24} shows an example of the dynamics of the magnetization for $N=9$ in the time crystal phase, compared to the exact simulations (orange dashed line). They match perfectly. [Panel (b)] compares the entanglement entropy obtained with finite time resolutions and the exact simulation. In this case, we observe a strong dependence on the bin size of the steady-state value.

\begin{figure}
    \centering
    \includegraphics[width=0.8\linewidth]{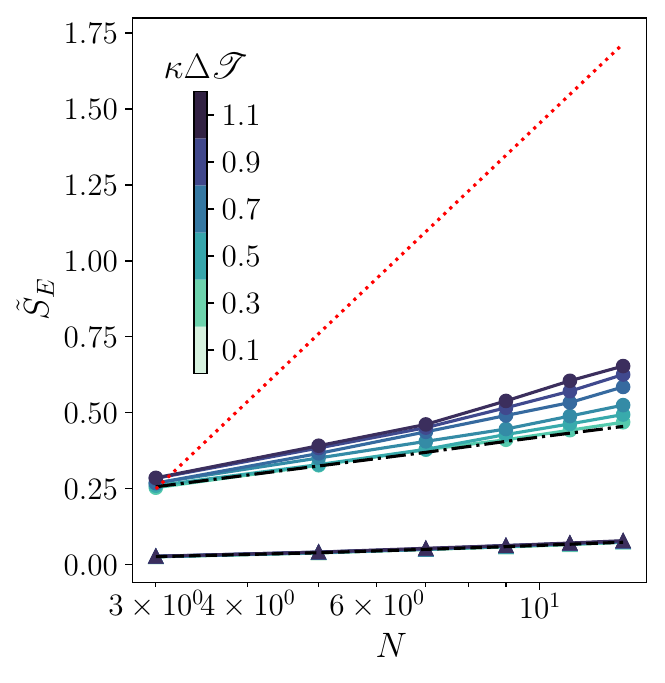}
    \caption{Steady state value of the entanglement entropy with respect to system size for several bin widths. $\blacktriangle$ markers correspond to the stationary phase, the $\bullet$ markers correspond to the time-crystal phase. The exact results for the stationary and time crystal steady-state $S_E$ are reported respectively with a dashed and dash-dot line. The red dot line shows $\log N$ for reference. The average is obtained from 1000 mixed-state trajectories.} 
    \label{figAnna_app3}
\end{figure}

In Fig.~\ref{figAnna_app3} we show the steady-state value of the entanglement entropy in the stationary phase (with markers $\blacktriangle$) and in the time-crystal phase (with $\bullet$ markers). The stationary phase has the property of having a low number of photons in the steady state, resulting in having few jumps along trajectories. This causes the steady-state value of $\tilde S_E$ to be robust to variations of the binning size. We always obtain results that are consistent with the exact simulations (black dashed line). The time-crystal phase, exhibiting a higher number of photons and jumps, is strongly affected by the change in the bins' size and deviates from the exact result (black dash-dot line). Also in this case, the scaling is given by $(\log N)^\alpha$ with $\alpha<1$. The power $\alpha$ increases for increasing bin sizes, as the entanglement entropy in these cases also accounts for classical correlations. Notice however that a resolution $\Delta \mathscr T \sim 1/\kappa$ yields a scaling qualitatively different from $\log N$, which corresponds to the scaling of the (entanglement) entropy given by the Lindblad equation (red dotted line). Thus, we can say that for $\Delta \mathscr T \sim 1/\kappa$ the entropy is mostly governed by quantum correlations, and the entanglement transition can be witnessed by the experimental device.

As a closing remark, we notice that our approximation of the master equation is reliable as long as the typical waiting time $\propto (\kappa\expval{\hat a^\dagger \hat a})^{-1}$, which scales at most as $N^{-1}$, is smaller than the bin size $\Delta \mathscr T$. To take into account larger sizes with a smaller waiting time one should allow for the occurrence of more than one jumps in the master equation. For example, allowing for the presence of two jumps one would have the following modified master equation:  the non-detection state \eqref{eq_binme_nj} occurs with probability $p_0=1-\kappa\Delta \mathscr T\expval{(\hat a^\dag\hat a)^2}$; the detection state, happening with probability $p_{1,2}=p_1+p_2=\kappa\Delta \mathscr T\expval{(\hat a^\dag\hat a)^2}$ is instead modified to
\begin{eqnarray}
     && \hat\rho^{1,2 \rm{J}}_\gamma = \frac{p_1}{p_{1,2}} \hat\rho^{1\rm{J}}_\gamma+\frac{p_2}{p_{1,2}}\hat\rho_\gamma^{2\rm{J}}
\end{eqnarray}
with $\hat\rho_\gamma^{1\rm{J}}$ as in eq. \eqref{eq_binme_j} and $\hat\rho_\gamma^{2\rm{J}}$ being the average over all the trajectories having two jumps inside the bin. The probabilities in the above expression are defined as $p_1=\kappa\Delta \mathscr T\expval{\hat a^\dag\hat a}$ (probability of having one jump) and as $p_2=\kappa\Delta \mathscr T(\expval{(\hat a^\dag\hat a)^2}-\expval{(\hat a^\dag\hat a)})$ (probability of having two jumps).

\section{Gaussian trajectory approximation}\label{sec:gaussian-bhd}

In our study of the Bose-Hubbard dimer, we further perform the \textit{Gaussian approximation} on top of the quantum-jump unraveling defined in Eqs.~\eqref{eqAnna:nojump} and~\eqref{eqAnna:jump}, since the exact simulation of the system is numerically challenging especially when approaching the thermodynamic limit ($N\to\infty$). This method was introduced in Refs.~\cite{verstraelen2018gaussian,verstraelenGaussianTrajectoryApproach2020a} and has been demonstrated to work well in semiclassical bosonic systems. The Gaussian approximation consists of assuming the state along any unraveled trajectory to be a Gaussian state, which is essentially a variational ansatz completely parametrized by its first and second moments, i.\,e., $\langle\aaa_i\rangle$, $\langle\aaa_i\aaa_j\rangle$, and $\langle\daaa_i\aaa_j\rangle$. In practice, we define and evolve the following quantities:
\bea
    R&\equiv \norm{\psitilde}^2\,,\\
    \alpha_i&\equiv\langle\aaa_i\rangle\,,\\
    u_{ij}&\equiv \langle \aaa_i\aaa_j\rangle-\alpha_i\alpha_j\,,\\
    v_{ij}&\equiv \langle\daaa_i\aaa_j\rangle-\alpha_i^*\alpha_j\,,
\eea
where $\psitilde$ denotes the unnormalized wavefunction. In the quantum-jump unraveling, the state between jumps evolves under the effective non-hermitian Hamiltonian $\Hhat-\rmi\frac{\kappa}{2}\sum_i\daaa_i\aaa_i$ according to Eq.~\eqref{eqAnna:nojump}. The wavefunction norm is therefore not conserved and evolves as
\bea\label{eq:dRdt-bh}
    \dfrac{\d R}{\d t} = -\kappa\sum_i\langle\daaa_i\aaa_i\rangle R\,.
\eea
For a time-independent observable $\ohat$, its normalized expectation value propagates under this non-unitary evolution as
\bea\label{eq:dodt-bh}
    \dfrac{\d\langle\ohat\rangle}{\d t} &= \rmi\left\langle[\Hhat,\ohat]\right\rangle \\
    & + \kappa\sum_i\left( \langle\daaa_i\aaa_i \rangle\langle\ohat\rangle - \dfrac{1}{2}\langle\daaa_i\aaa_i\ohat\rangle-\dfrac{1}{2}\langle\ohat\daaa_i\aaa_i\rangle\right)\,.
\eea

This evolution is interrupted by jumps occurring at random times, where the wavefunction updates as $\psitilde\to \aaa_i\psitilde/\sqrt{||\aaa_i\psitilde||^2}$ with probability $\pscr_i(\dt)=\kappa\langle\daaa_i\aaa_i\rangle\dt$ for each infinitesimal interval $\dt$. This translates into the update rule of the expectation
\bea\label{eq:update-o-bh}
    \langle\ohat\rangle\longrightarrow\dfrac{\langle\daaa_i\ohat\aaa_i\rangle}{\langle\daaa_i\aaa_i\rangle}\,,\quad R\longrightarrow 1\,.
\eea

Under the Gaussian approximation, we can express any expectation involving products of more than two bosonic operators in terms of the Gaussian variational parameters 
$\alpha_i$, $u_{ij}$ and $v_{ij}$, by repeated applying the Wick's theorem.
A closed set of equations of these quantities can therefore be obtained from Eqs.~\eqref{eq:dRdt-bh}-\eqref{eq:update-o-bh}. The detailed equations of motions are given below.

The advantage of the Gaussian trajectory approximation compared to phase-space trajectory approaches (such as the truncated Wigner approximation~\cite{Vogel1989}) is that each Gaussian trajectory fully characterizes the evolution of a physical state represented by the Gaussian parameters, which allows access to any physical quantity on the level of single trajectories. For example, the entanglement entropy between the two bosonic modes can be calculated as~\cite{serafiniSymplecticInvariantsEntropic2003}:
\bea
    S_E &=\left(\mu + \dfrac{1}{2}\right) \log\left(\mu + \dfrac{1}{2}\right) - \left(\mu - \dfrac{1}{2}\right)  \log\left(\mu - \dfrac{1}{2}\right)\,,\\
    \mu &= \sqrt{|\Vec{\sigma}|}\,,
\eea    
where $\sigma_{ij}\equiv\frac{1}{2}\langle  \hat{r}_i \hat{r}_j +  \hat{r}_j \hat{r}_i \rangle - \langle\hat{r}_i\rangle\langle\hat{r}_j\rangle$ is the 
covariance matrix of one of the modes, with the standard definition $\hat{\Vec{r}}\equiv(\hat{x},\hat{p})$, $\hat{x}=(\daaa+\aaa)/\sqrt{2}$, $\hat{p}=\rmi(\daaa-\aaa)/\sqrt{2}$, where $\aaa$ denotes the bosonic annihilation operator of the mode in question. 
\onecolumngrid

\subsection{Detailed expressions for the equations of motions of the Bose-Hubbard dimer in the Gaussian approximation}\label{app:bhd-eom}

We give below the equations of motion of the quantum jump unraveling of the Bose-Hubbard model on a generic lattice, defined by the following Hamiltonian and master equation:
\bea
    \Hhat &= \sum_i \left(-\Delta_i\daaa_{i}\aaa_{i} + \frac{U_{i}}{2}\daaas_{i}\aaas_{i} + F_{i}\daaa_{i} + F^*_{i}\aaa_{i}\right) - \frac{1}{2}\sum_{i,j}\left(J_{ij}\daaa_{i}\aaa_{j} + J^*_{ij}\daaa_{j}\aaa_{i}\right)\,,\\
    \dfrac{\d\rhohat}{\d t}&=-\rmi[\Hhat,\rhohat]+\sum_j \kappa_j\(\aaa_j\rhohat\daaa_j-\dfrac{1}{2}\{\daaa_j\aaa_j,\rhohat\}\)\,,
\eea
where we assume the on-site detuning $\Delta_i$, nonlinearity $U_i$, drive $F_i$ and dissipation $\kappa_i$ to be generic and site-dependent. The connectivity of the lattice is specified by the coupling matrix $J_{ij}$. Within the Gaussian approximation and the quantum jump unraveling,
the deterministic evolution between jumps is given by

\bea\label{eq:pc-determ}
    \dfrac{\d R}{\d t} =&~ -\d t\sum_j \kappa_j N_j R \,,\\
    \dfrac{\d \alpha_n}{\dt}=&~ \(-\dfrac{\kappa_n}{2}+\rmi \Delta_n\)\alpha_n+\rmi\sum_j J_{nj}\alpha_j \\
    &- \rmi U_n\(\abss{\alpha_n}\alpha_n+2\alpha_n v_{nn} + \calpha_n u_{nn}\)   - \rmi F_n\\
    &-\sum_j\kappa_j A^{(\alpha)}_{nj}\,,\\
\dfrac{\d u_{nm}}{\dt}=&~\[-\dfrac{\kappa_m+\kappa_n}{2}+\rmi(\Delta_n+\Delta_m)\]u_{nm}+\rmi \(\sum_{n'}J_{nn'}u_{n'm}+\sum_{m'}J_{mm'}u_{nm'}\)\\
& - \rmi U_n\[v_{nm}\(\alpha^2_n + u_{nn}\) + 2 u_{nm} \(\abss{\alpha_n}+v_{nn}\)\]\\
& - \rmi U_m\[v_{mn} \(\alpha_m^2 + u_{mm}\) + 2 u_{mn} \(\abss{\alpha_m}+v_{mm}\) + \delta_n^m \(\alpha_m^2 + u_{mm}\)\]\\
& - \sum_j \kappa_j A^{(u)}_{nmj}\,,\\
    \dfrac{\d v_{nm}}{\dt} =&~ \[ \rmi \( \Delta_m - \Delta_n \) - \dfrac{\kappa_m + \kappa_n }{2}  \]v_{nm} + \rmi \( \sum_{m'}  J_{mm'} v_{nm'} - \sum_{n'}  J_{n'n} v_{n'm} \)\\
    & + \rmi  U_n \[ 2 v_{nm} \(\abss{\alpha_n} + v_{nn} \) + u_{nm} \( \calphas_n + u_{nn}^\ast \) \] \\ 
    & - \rmi  U_m \[ 2 v_{nm} \(\abss{\alpha_m} + v_{mm} \) + u_{nm}^\ast \(\alpha_m^2 + u_{mm}\) \] \\
    & -\sum_j \kappa_j A^{(v)}_{nmj}\,, \\
\eea

where
\bea
    N_j =&~ \abss{\alpha_j} + v_{jj}\,,\\
    A^{(\alpha)}_{nj} =&~ \alpha_j v_{jn} + \calpha_j u_{jn}\,,  \\
    A^{(u)}_{nmj} =&~  v_{jn} u_{jm} + v_{jm} u_{jn}\,,\\
    A^{(v)}_{nmj} =&~  v_{jm}v_{nj} + u_{jn}^\ast u_{mj}\,. \\
\eea

When a jump associated to the jump operator $\aaa_j$ occurs, the variational parameters update as follows:

\bea
    R ~ \rightarrow & ~ 1\,,\\ 
   \alpha_n ~ \rightarrow & ~ \alpha_n + \dfrac{ A^{(\alpha)}_{nj} }{ N_j } \,,\\
   u_{nm} ~ \rightarrow & ~ u_{nm} + \dfrac{ A^{(u)}_{nmj} }{N_j} - \dfrac{  A^{(\alpha)}_{nj}  A^{(\alpha)}_{mj} }{ \( N_j \)^2 } \,,\\
   v_{nm} ~ \rightarrow  & ~ v_{nm} + \dfrac{ A^{(v)}_{nmj}  }{N_j} - \dfrac{ \( A^{(\alpha)\ast}_{nj} \) A^{(\alpha)}_{mj} }{ \( N_j \)^2 }\, ,
\eea
and the probability for the jump to occur at each time step $\dt$ is 
\bea
    P_j\(dt\) = \kappa_j\mean{\daaa_j\aaa_j}\dt = \kappa_j N_j \dt\,.
\eea

\twocolumngrid
\section{More details on experimental implementations}\label{app:ExpImp}

In this Appendix we provide some numbers for the cold atoms experimental implementation of the Bose-Hubbard dimer described in Section~\ref{DBH}. To begin with, from comparison with other atoms with similar level structures, the ground state splittings caused by the tensor light shifts can be as large as of $2\pi\times 1-10$~MHz [see Fig.~\ref{figEric:Implementation}(c)]. The couplings induced by the magnetic field and/or Raman transitions should be such that their associated Rabi frequencies are no larger than $~\sim 2\pi\times 100$~kHz to avoid coupling $\ket{1}$ and $\ket{2}$ to the rest of nuclear sublevels.

As described in the main text, generating interactions requires dispersively coupling the $^1 S_0\leftrightarrow \!^3P_1$ optical transition to a far-detuned cavity mode (with detuning $\Delta_{ac}$) and driving the cavity with $\hat{z}$ polarized light that is moderately detuned from the $^1S_0\leftrightarrow ^3P_1$ transitions (with detuning $\Delta_{da}$ and Rabi frequency $\Omega_d$). This gives rise to the four-photon process depicted in Fig. The associated $U$ is approximately given (up to factors of order $1$ related to Clebsch-Gordan coefficients) by~\cite{Chu2023}
\begin{equation}
    U\approx \left(\frac{C\kappa}{\Delta_{ac}}\right)\left(\frac{\Omega_d^2}{\Delta_{da}^2}\right)\gamma,
\end{equation}
where $C$ and $\kappa$ are the cooperativity and FWHM linewidth of the cavity, respectively, and $\gamma$ is the decay rate out of the excited $^3P_1$~($F=3/2,m_F=\pm 1/2$) manifold. To guarantee that unitary dynamics is not affected by $\gamma$, which induces incoherent ground state processes at a rate $\gamma (\Omega_d/\Delta_{da})^2$, we need $C\kappa/\Delta_{ca}\gg 1$, which can be achieved by using a cavity with large enough cooperativity. Noting that $\gamma^{-1}\approx 1~\mu s$~\cite{Hoffrichter2016}, we can guarantee that $U\lesssim 2\pi \times 100$kHz by using adequate $\Omega$ and $\Delta_{da}$.

\begin{figure}
    \centering
    \includegraphics[width=0.75\linewidth]{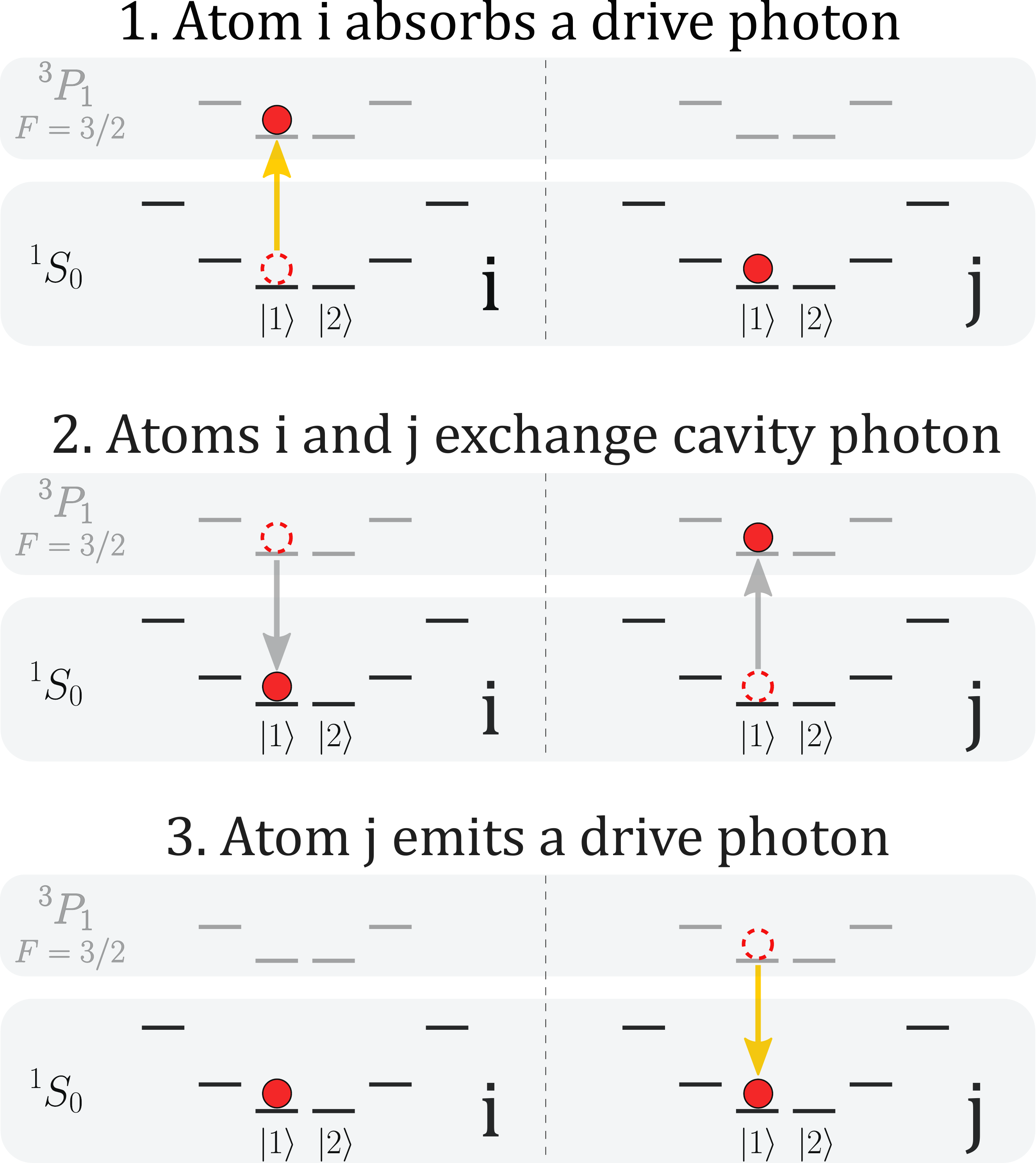}
    \caption{Four photon process between atoms $i$ and $j$. First a drive photon (yellow) is absorbed by an atom $i$. Then, atom $i$ emits a cavity photon with $\pi$ polarization, and this cavity photon is absorbed by atom $j$. Finally, atom $j$ emits a photon stimulated by the drive. The process depicted in the figure leads to a density-density terms of the form $(\hat{a}_1^\dagger)^2\hat{a}_1^2$, and there is a similar process for atoms in $\ket{2}$.}
    \label{figEric:appG}
\end{figure}

\end{document}